\documentclass[twocolumn,pre,superscriptaddress,floatfix,amsmath,amssymb,aps]{revtex4-2}
\pdfoutput=1

\usepackage[utf8]{inputenc}
\usepackage{natbib}
\usepackage{array}
\usepackage{booktabs}
\usepackage{graphicx}
\usepackage{paralist}
\usepackage{xcolor}
\usepackage{color,soul}
\usepackage[colorlinks=true, allcolors=blue]{hyperref}
\usepackage{mathtools}
\usepackage{amsmath}
\usepackage{dcolumn}
\usepackage{amssymb}
\usepackage{gensymb}
\usepackage{mathdesign}
\usepackage{siunitx}
\usepackage{accents}
\usepackage{bm}
\usepackage{dcolumn}
\usepackage{float}

\begin{document}

\title{Reverse segregation and self-organization in inclined chute flows of bidisperse granular mixtures}

\author{Joseph M. Monti}
\affiliation{Sandia National Laboratories, Albuquerque, NM 87185, USA}
\author{Joel T. Clemmer}
\affiliation{Sandia National Laboratories, Albuquerque, NM 87185, USA}
\author{Ishan Srivastava}
\affiliation{Center for Computational Sciences and Engineering, Lawrence Berkeley National Laboratory, Berkeley, California 94720, USA}
\author{Leonardo E. Silbert}
\affiliation{School of Math, Science, and Engineering, Central New Mexico Community College, Albuquerque, New Mexico 87106, USA}
\author{Gary S. Grest}
\affiliation{Sandia National Laboratories, Albuquerque, NM 87185, USA}
\author{Jeremy B. Lechman}
\affiliation{Sandia National Laboratories, Albuquerque, NM 87185, USA}
\date{\today}

\begin{abstract}
In the usual segregation scenario for stable inclined chute flows of bidisperse mixtures of fine and coarse spherical particles, coarse particles rise toward the free surface, forming a coarse-rich region atop the flowing pile.
Beyond a threshold coarse-to-fine diameter ratio of approximately 4, conversely, the weight of the coarse particles exceeds the segregation driving forces, causing individual coarse particles to sink within the pile and producing a reversed segregation state.
However, an understanding of the collective evolution of the pile structure is still lacking when the particle diameter ratio exceeds 4 {\textit{and}} the coarse particle mass fraction is appreciable.
To explore this broadly bidisperse limit, we perform discrete element method simulations considering mean particle diameter ratios of up to 8 and coarse particle mass fractions spanning 0.1 to 0.9. 
The steady-state flow profiles reveal several intriguing behaviors that depend on the diameter ratio and mass fraction. 
These include a previously identified transition from usual to reverse segregation and a newfound tendency to self-organize into alternating coarse- and fine-rich particle layers stacked along the shear gradient direction, with layer thickness dictated by the coarse particle diameter.
A fuller understanding of segregation at this scale could pave the way for enhanced mixing or demixing techniques at the commercial scale.
\end{abstract}

\maketitle

\section{Introduction}
Granular segregation induced by particle size dispersity is a commonplace and often-researched occurrence in sheared or gravity-driven flows~\cite{gray2018,umbanhowar2019} and in rotating~\cite{seiden2011} and vibrated systems~\cite{kudrolli2004}.
In the context of inclined chute flows of bidisperse mixtures of dry, spherical granular particles, the phenomenon evokes an image of coarse particles rising through a pile of flowing fine particles of similar material density to the free surface and remaining there until the conditions change or flow ceases~\cite{savage1988}.
However, the mechanisms that drive segregation must compete with the coarse particle weight, which grows as the particle diameter cubed.
Segregation mechanisms that scale more slowly with diameter---for example, any that scale like the diffusion coefficient (which is proportional to the volume-weighted mean particle diameter squared~\mbox{\cite{barker2021,trewhela2021,liu2023}})---are outpaced at large diameter ratios.
Intuition therefore suggests that a threshold coarse-to-fine particle diameter ratio exists beyond which the gravitational force on any single coarse particle exceeds the effective segregation force, producing a situation in which the particle would sink rather than rise---similar in spirit, if not strictly in mechanism, to the reverse Brazil nut effect (BNE)~\cite{shinbrot1998,hong2001,jenkins2002,breu2003,garzo2009}. 
Traditionally, a transition from BNE to reverse BNE is observed in granular beds vibrated at sufficient amplitude to preferentially mobilize fine particles, allowing coarse particles to descend into lower, vacant positions~\mbox{\cite{breu2003}}.
The vibrated bed transition is primarily governed by the particle diameter and density ratios~\mbox{\cite{hong2001,jenkins2002,breu2003,garzo2009}}, which was also shown to be the case for dense sheared flows with modest diameter ratios~\mbox{\cite{tunuguntla2014}}.

Discrete element method (DEM) simulations have been used to quantify the rise-or-sink threshold for individual coarse particle intruders embedded in dense sheared or free surface flows under gravity~\cite{guillard2016,vanDerVaart2018,jing2020,jing2021}.
For free surface inclined chute flows of otherwise similar spherical particles with a coarse-to-fine particle diameter ratio denoted $\alpha$, the threshold $\alpha$ at which the effective segregation force equals the weight is approximately 4~\cite{jing2020,jing2021}.
Thus, while segregation toward the pile surface occurs in bidisperse mixtures of relatively modest $\alpha\lesssim 3$, mixtures with large enough $\alpha$ should seemingly undergo the reverse dynamics, wherein coarse particles descend within the pile.
~\citet{thomas2000} experimentally identified such a reversal of segregation behavior (see also~\cite{felix2004,thomas2018}).
For inclined chute flows of bidisperse mixtures of spherical glass beads with coarse particle mass fractions of 0.1 and 0.9, ``usual'' segregation transpires when the diameter of the less abundant species by mass fraction is smaller than roughly 4 times the diameter of the more abundant species, in alignment with the DEM intruder simulations~\cite{jing2020,jing2021}.
In contrast, ``reverse'' segregation, i.e., fine particles covering the pile surface, prevails once the diameter of the more abundant species by mass fraction is larger than that of the less abundant species by a factor of roughly $5$.  
\citet{thomas2000} further reported that in reverse segregation scenarios, the coarse particles were positioned at intermediate heights in the pile---none were located at the top surface, nor had any sunk to the base.
Most notably, borderline behavior was identified for the diameter ratio of 4.3, which yielded coarse particles located throughout the pile and at the top surface.
This measured threshold was recapitulated in more recent DEM simulations of inclined chute flow by~\citet{thomas2018}, which confirmed the settling of sufficiently large coarse tracer particles deep into the pile.

The importance of the particle diameter ratio in the transition between usual and reverse segregation for individual or tracer-level coarse particle mass fractions begs the question of what the effects of jointly increasing the size and relative abundance of coarse particles are on the segregation character of the flowing pile.
Moreover, it is unclear whether the corresponding flow still adheres to the Bagnold rheology that has been shown to be valid for monodisperse inclined chute flows~\cite{bagnold1954,silbert2001,midi2004}, as has been contended for bidisperse mixtures with diameter ratios of approximately 2 or less~\cite{rognon2007,tripathi2011,staron2014,barker2021,singh2024}.
To probe these questions, we performed DEM simulations to study inclined chute flow of bidisperse mixtures with mean particle diameter ratios of 2, 4, 6, and 8 and with coarse particle mass fractions in the range 0.1 to 0.9.
This range of parameters contains systems that can be compared with previous modeling and experimental results and extends into the regime where free sifting occurs~\cite{gao2023}.
Of particular interest is the local particle arrangement and packing density as the composition of the pile changes, as bidispersity has substantial ramifications for the compactness of jammed~\cite{srivastava2021a} and flowing systems~\cite{golick2009,tripathi2011}. 
Consequently, this article focuses on establishing the flow phenomenology for these broadly size-bidisperse systems.
Detailed mechanistic explanations for the observed behaviors are left to future work.

The organization of the article is as follows.
Section~\ref{sec:methods} describes the DEM simulation methodology.
In Sec.~\ref{sec:segbe}, we report steady-state profiles for local coarse particle concentrations and particle packing densities as a function of pile composition.
Section~\ref{sec:trans} explores the transition between usual and reverse segregation behaviors, and Sec.~\ref{sec:velocity} shows steady-state profiles of flow speed and associated shear rates.
Section~\ref{sec:conc} briefly summarizes our main results.

\section{Methods}
\label{sec:methods}
\begin{figure}
\centering
\includegraphics[width=0.35\textwidth]{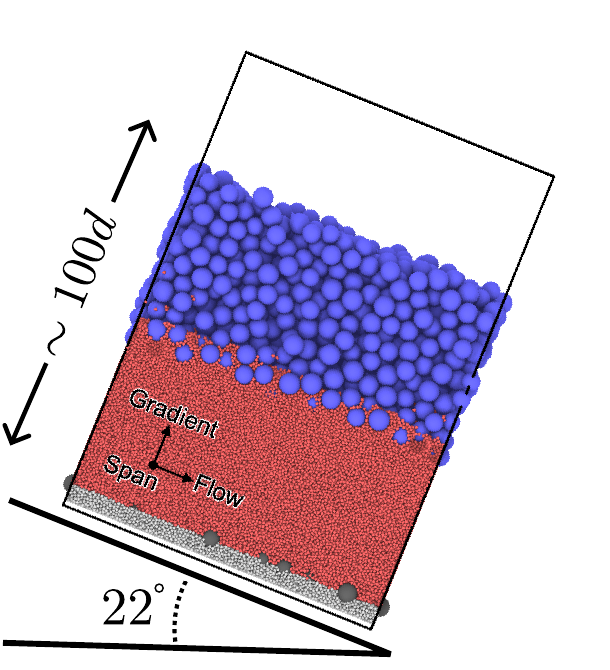}
\caption{
Snapshot of the initial configuration for $\alpha = 6$ and $f_c = 0.5$.
Mobile coarse and fine particles are shown in blue and red, respectively.
The frozen base also consists of fine (light gray) and coarse particles (dark gray).
}
\label{fig:fig1}
\end{figure}

Three dimensional DEM inclined chute flow simulations of bidisperse mixtures of spherical particles were performed using the GRANULAR package in LAMMPS~\cite{plimpton1995,thompson2022}, which efficiently simulates practically arbitrary particle diameter ratios~\cite{monti2022a}.
Each simulation consisted of fine particles with mean diameter $d$ and coarse particles with mean diameter $\alpha d$, with $\alpha = 2, 4, 6, $ and 8.
Individual particle diameters of both species were uniformly distributed within $\pm 10\%$ of the species mean.
All particles were composed of the same generic material with intrinsic mass density set to unity, so that the mass scale $m$ of a typical fine particle is equal to $m = \pi d^3/6$.
The acceleration due to gravity, $g$, is also set to unity.
The characteristic simulation scales are therefore~\cite{silbert2001}: time $\tau_0 = \sqrt{d/g}$, stiffness $k_0 = mg/d$, and velocity $v_0 = \sqrt{gd}$.
The inter-particle contact model closely followed~\citet{silbert2001} (specifically model ``L3'').
Particles interacted along the normal direction via damped linear springs with spring constant $k \approx 3.8\cdot 10^5 k_0$.
The damping coefficients were set to model a coefficient of restitution of 0.9.
Frictional interactions between particles with a sliding friction coefficient of $\mu_s = 0.5$ were modeled using a linear, displacement history-dependent mechanism with stiffness equal to $2k/7$.
The simulation time step was set to $5\cdot 10^{-5}\tau_0$ to adequately resolve collisions between pairs of fine particles.
Additional test simulations showed that the results were not sensitive to further reductions of the time step size.

Construction of the pile configurations followed standard DEM practice for inclined chute flows~\mbox{\cite{silbert2001,rognon2007,tripathi2011,staron2014,thomas2018,thompson2022,liu2023}}.
An example initial pile configuration is shown in Fig.~\ref{fig:fig1}.
The modeled systems were constructed as deep and narrow to elucidate the bulk internal structure of flowing piles as a function of the particle diameter distribution while minimizing secondary flows~\cite{forterre2001,dOrtona2020}.
In the free surface inclined flows considered, neither volume confinement nor overburden pressure were imposed~\mbox{\cite{golick2009,umbanhowar2019,jing2020,jing2021}}, thereby allowing the pile to shrink or swell as dictated by the pile composition and flow rate.
The simulation cell was periodic along the flow ($x$) and span ($y$) directions.
The gradient direction aligned with the $z$ axis.
For $\alpha = 2$ and 4, the flow dimension had length $50d$, which was doubled for $\alpha = 6$ and 8 to $100d$.
In all simulations, the length of the span dimension was $50d$.
Non-inclined initial configurations were constructed by depositing a block of coarse particles onto a block of fine particles to produce a combined resting thickness of approximately $100d$, with the relative block widths set to control $f_c$, the fraction of the total particle mass contributed by coarse particles: $f_c = N_c\alpha^3/\left(N_f + N_c\alpha^3\right)$, where $N_f$ and $N_c$ are the number of fine and coarse particles, respectively.
While the coarse species was initially arrayed across the pile surface to emphasize ensuing evolution during flow, we also confirmed that our results were reproducible with initially homogeneous mixtures.
The base, located at $z = 0$, was composed of frozen particles to minimize slip~\mbox{\cite{silbert2002,goujon2003,kumaran2012,weinhart2012,jiang2023,liu2023,guo2025}}.
In addition to frozen fine particles that were taken from a previously jammed disordered packing, frozen coarse particles covering $\approx 28\%$ of the basal surface area were inserted near $z=0$ to mitigate coarse particle basal slip~\mbox{\cite{goujon2003,guo2025}}.

\begin{figure*}
\centering
\includegraphics[width=0.9\textwidth]{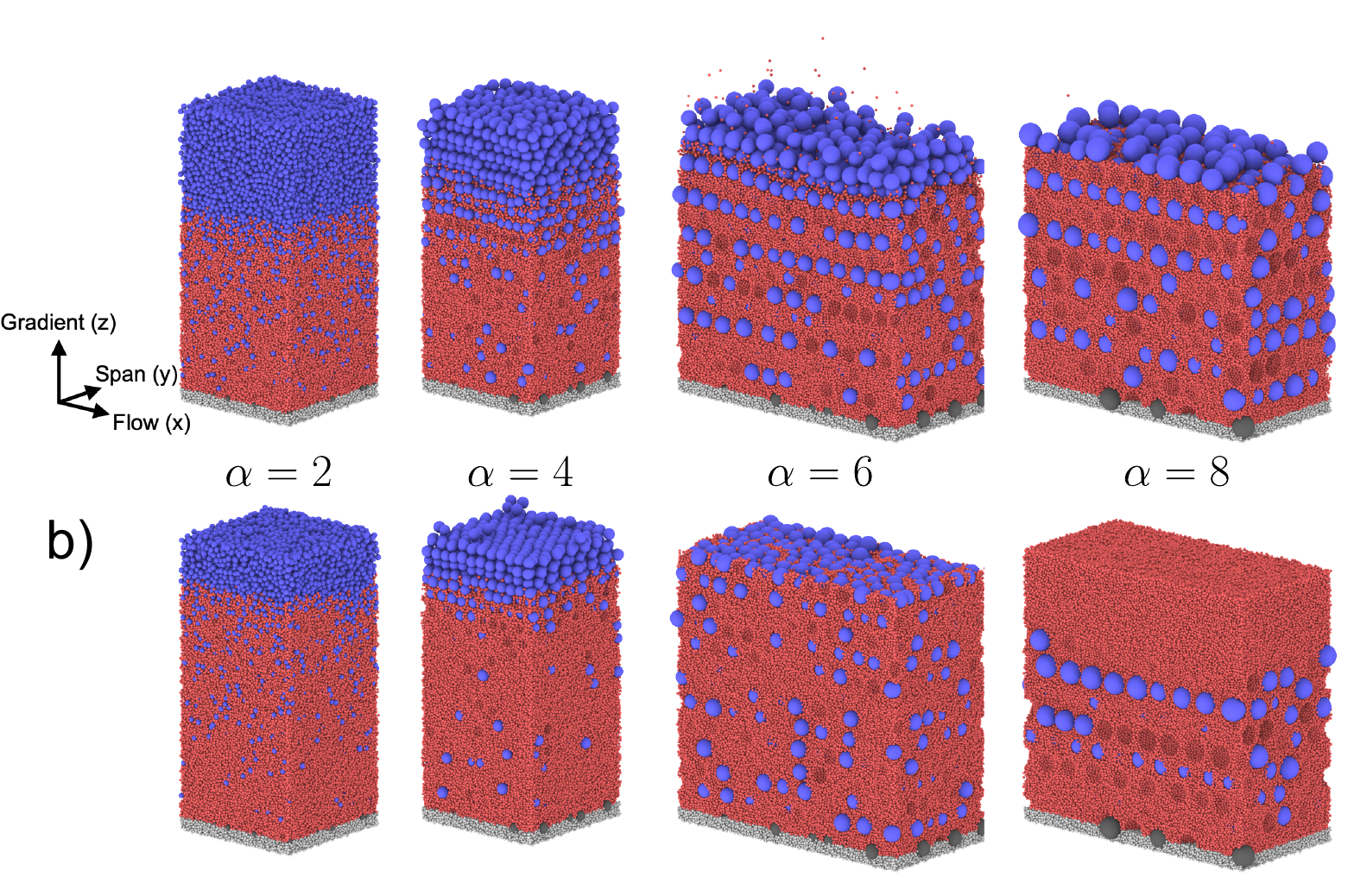}
\caption{
Flowing configurations obtained at $t = 2\cdot 10^4\tau_0$ for a) $f_c = 0.5$ and b) $f_c = 0.3$ for the indicated $\alpha$.
The respective flow and span base dimensions are $50d$ by $50d$ for $\alpha = 2$ and 4 and $100d$ by $50d$ for $\alpha = 6$ and 8.
The images were rendered in OVITO~\cite{ovito}.
}
\label{fig:fig2}
\end{figure*}

To initiate flow, the base was inclined instantaneously at $t = 0$ to an angle of $\theta = 22\degree$, a typical value for inclined chute flow simulations of frictional particles~\mbox{\cite{silbert2001,tripathi2011,kumaran2012,staron2014}}.
Our results are consistent across a range of higher incline angles; however, given the magnitude of the parameter space explored in this work, other angles were not investigated systematically.
Simulations were performed over a time interval lasting at least $2\cdot 10^4\tau_0$, which was usually a sufficient duration to reach steady state with respect to evolution of the pile.
The cumulative flow distance was $\mathcal{O}\left(10^4d\right)$, which is a considerable distance for laboratory-scale granular materials~\cite{thomas2000,forterre2001,dOrtona2020}.

Flow profiles of interest, including coarse particle concentration, packing density, and flow velocity and shear rates, were computed by defining bins of size $10^{-2}d$ along the $z$ axis~\mbox{\cite{kumaran2013,jing2021,liu2023}}.
For each bin, the properties of intersected particles were tabulated by weighting by particle cross-sectional area in the associated horizontal plane, constituting an average taken over the span and flow directions.
Velocity variances were calculated using a larger bin size to reduce noise.
Flowing configurations were saved every $2\cdot 10^2\tau_0$ and, in most cases, results were averaged over the final $2\cdot 10^3\tau_0$ to reduce scatter in the data.
Finally, the instantaneous surface height of the flowing pile was computed by determining the position of the highest bin with a packing density of at least 0.35.
Long time flowing heights for each of the simulated systems are shown in the Supplemental Materials.

\begin{figure*}
\centering
\includegraphics[width=0.8\textwidth]{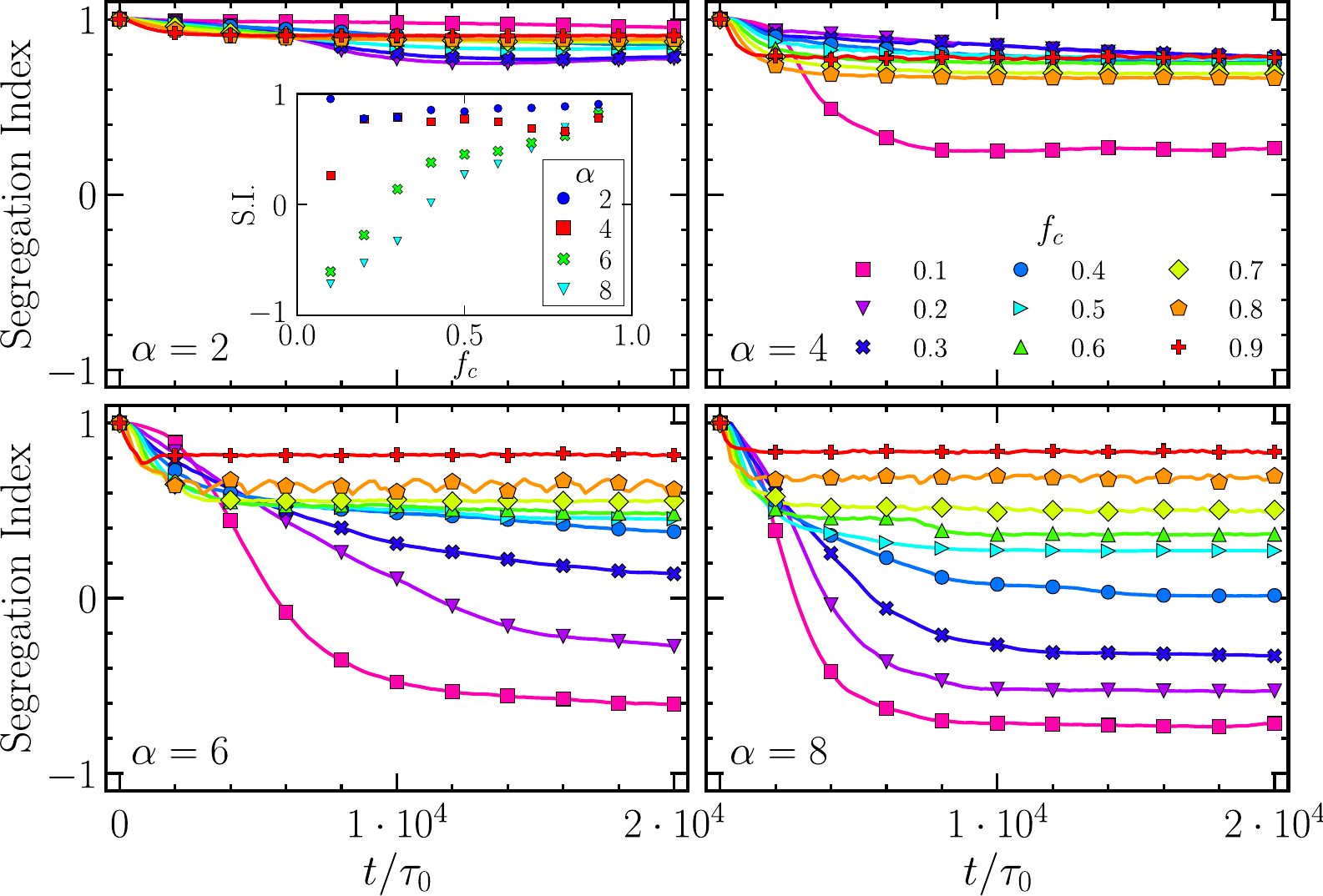}
\caption{
Evolution of the segregation index with time for four values of $\alpha$ and the nine values of $f_c$ indicated in the legend of the $\alpha = 4$ panel.
Data are normalized to unity at $t = 0$.
Markers are shown at fixed intervals to differentiate the curves.
The inset of the $\alpha = 2$ panel shows the segregation index obtained at $t = 2\cdot 10^4\tau_0$ for each $\alpha$ and $f_c$ combination.
}
\label{fig:fig3}
\end{figure*}

\section{Results}
\label{sec:results}

~\citet{thomas2000} provided an experimental phase diagram delineating the boundary between usual and reverse segregation regimes for inclined chute flows of bidisperse mixtures with coarse particle mass fractions of $f_c = 0.1$ and 0.9.
The two regimes were chiefly differentiated by the surface composition of the pile.
Here, we study the steady-state behavior of the full $f_c$ range for $\alpha \leq 8$ to systematically explore the collective segregation of bidisperse mixtures with large particle diameter ratios, for which the steady-state surface and interior structure are generally unknown.

\subsection{Description of Segregation Behavior}
\label{sec:segbe}

Figure~\ref{fig:fig2} shows side by side configurations of simulated inclined chute flows at an incline angle of $\theta = 22\degree$ for $\alpha = 2, 4, 6, $ and 8.
In each case, the elapsed flow time is $\Delta t = 2\cdot 10^4\tau_0$.
The configurations in Fig.~\ref{fig:fig2}a) correspond to $f_c = 0.5$ and Fig.~\ref{fig:fig2}b) shows $f_c = 0.3$.
The images in Fig.~\ref{fig:fig2}a) all show usual segregation behavior strictly in that coarse particles cover the pile surface.
Almost no fines reside within the upper region of the pile for any $\alpha$; although some fines are temporarily airborne [e.g., in the $\alpha = 6$ image in Fig.~\ref{fig:fig2}a)], they readily percolate through the coarse particles upon returning to the pile~\cite{gao2023}.
The surface coarse particle domain varies in thickness from many coarse particle diameters for $\alpha = 2$ to the thickness of a single particle layer for $\alpha = 8$.
Below, the fine-dominant region never lacks coarse particles.
For $\alpha = 2$ and 4 in Fig.~\ref{fig:fig2}a), the concentration of coarse particles appears to decrease steadily away from the interface between the coarse-dominant and fine-dominant regions, whereas coarse particles occupy the entire pile for $\alpha = 6$ and 8.

The images for $f_c = 0.3$ shown in Fig.~\ref{fig:fig2}b) have both similarities to and differences from those for $f_c = 0.5$.
The $\alpha = 2$ and 4 configurations continue to exhibit usual segregation, but their coarse-dominant surface regions are reduced in thickness compared to the higher $f_c$ images. 
Conversely, the surface of the $\alpha = 6$ configuration is plainly mixed, and the $\alpha = 8$ configuration has undergone reverse segregation, since its surface is entirely composed of fines.
The sequence of images in Fig.~\ref{fig:fig2}b) suggests that $\alpha = 6$ and $f_c = 0.3$ produces a state intermediate between usual and reverse segregation, analogous to the transitional $\alpha = 4.3$ and $f_c = 0.1$ configuration that was identified by~\citet{thomas2000}.

For $\alpha = 6$ and 8 in Fig.~\ref{fig:fig2}, the images feature a distinctive, alternating, planar layered structure.
Similar layering occurs for $\alpha = 4$ and is most apparent in the vicinity of the border between coarse- and fine-dominant regions, with additional ordering in the unmixed coarse particle region at the top.
No layering was observed for $\alpha = 2$.
Flow-induced self-organization into horizontal layers is a recurring motif throughout this work and reflects a behavior that we found to be consistently robust with respect to simulation methodology.
To our knowledge, horizontal planar layers that extend throughout the pile in free surface inclined chute flows have not been unambiguously identified in prior work.
~\citet{thomas2018} noted that bands of coarse particles formed near the base for $f_c = 0.1$ under strong reverse segregation, but this represents an ordering caused by coarse particle proximity to the base.
Arguably,~\citet{rognon2007} observed layer formation in two dimensions (see Fig. 2 of that work, for instance), but no specific remarks upon this phenomenon were provided.
It is vital to point out that the alternating layered structure, while reminiscent of avalanche-induced layering in heaps~\cite{makse1997}, constitutes an entirely different behavior that manifests over a much smaller wavelength  $\sim (1+\alpha)d$.
In later sections, we analyze the properties of the layers and their formation at early times.

\begin{figure*}
\centering
\includegraphics[width=0.95\textwidth]{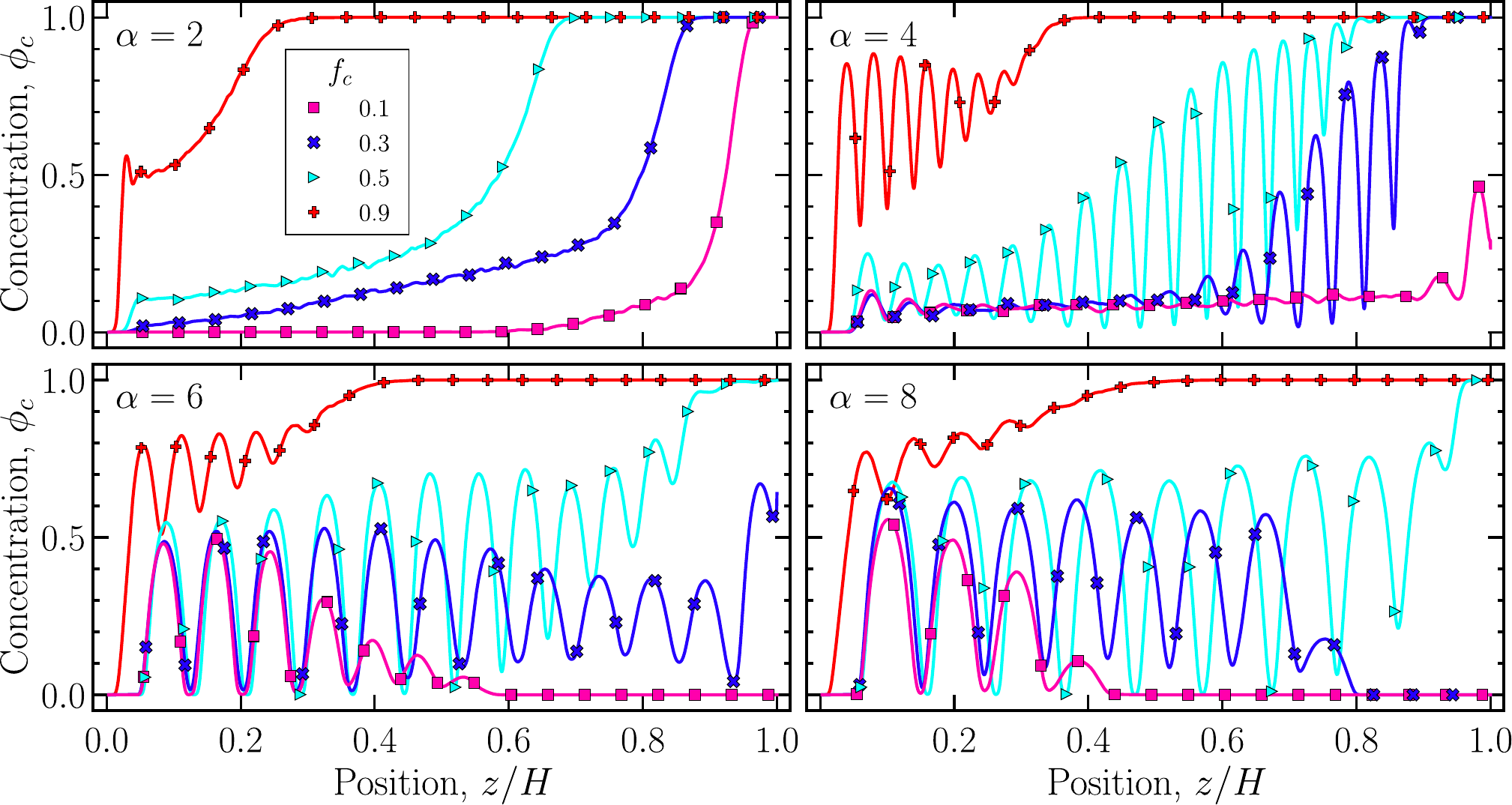}
\caption{
Steady state coarse particle concentrations for each $\alpha$ and the selected values of $f_c$ indicated in the legend of the $\alpha = 2$ panel.
Markers are shown at fixed intervals to differentiate the curves.
}
\label{fig:fig4}
\end{figure*}

Returning to the theme of characterizing segregation behavior, the degree of segregation can be quantified using the time-evolving index~\cite{dOrtona2020}
\begin{equation}
    \text{Segregation Index} = 2\frac{\textit{COM}_c-\textit{COM}_f}{H},
    \label{eq:si}
\end{equation}
in terms of the estimated flow height $H$ and the instantaneous center of mass ($COM$) coordinates of the coarse and fine species (subscripts $c$ and $f$) along the gradient direction.
The segregation index ranges from $+1$ to $-1$ for complete usual and reverse segregation, respectively.
Figure~\ref{fig:fig3} shows segregation index results obtained using Eq.~\eqref{eq:si} for all $\alpha$ and $f_c$ combinations.
The data are further normalized to $+1$ by their $t = 0$ values to compensate for slight differences of a few percent in the initial stacked configurations, which corresponded to usual segregation (see Fig.~\ref{fig:fig1}).

Figure~\ref{fig:fig3} shows that the segregation index drops below unity as flow begins, with most of the evolution occurring over the first $ 5\cdot 10^3\tau_0$.
For $\alpha = 2$, coarse particles segregate at the flow surface~\cite{thomas2000,tripathi2011,gray2018,umbanhowar2019,barker2021} despite the subsurface mixing visible in Fig.~\ref{fig:fig2} that corresponds to a 10--20\% reduction of the segregation index.
For $\alpha = 4$, results for $f_c > 0.1$ show similar behavior to $\alpha = 2$, but for $f_c = 0.1$, the segregation index plateaus at a value less than 0.5, indicating increased mixing.
Lastly, for $\alpha = 6$ and 8, Fig.~\ref{fig:fig3} shows that  the segregation index for most $f_c$ saturates at a distinguishable value that generally decreases with decreasing $f_c$, including to values approaching $-1$ that signify reverse segregation.
The trend is shown clearly in the inset in the upper left of Fig.~\ref{fig:fig3}, which plots the segregation index values obtained at $t = 2\cdot 10^4\tau_0$ against $f_c$.

The breadth of segregation index values for $\alpha = 4$ and higher shows that the degree of mixing strongly depends on $\alpha$ and $f_c$.
To resolve bulk mixing and other properties, we computed flow profiles as described in Sec.~\ref{sec:methods}.
Figure~\ref{fig:fig4} shows the steady-state values of the local coarse particle concentration, $\phi_c$, as a function of normalized position along the $z$ (gradient) axis for each $\alpha$ and selected values of $f_c$ (N.B., $\phi_f = 1-\phi_c$).
Usual and reverse segregation yield $\phi_c = 1$ and 0, respectively, at the pile surface ($z/H = 1$).
Results for $\alpha = 2$ show stable, homogeneous coarse particle regions at the surface that broaden with increasing $f_c$.
A mixed zone forms below the surface segregated region, and within, $\phi_c$ decreases smoothly with increasing depth but does not categorically vanish at the base for all $f_c$.

Focusing on trends of $\phi_c$ at the surface for $\alpha = 4$ and higher, results exhibit comparable features to $\alpha = 2$ with several key differences.
First, for $\alpha = 4$ and $f_c = 0.1$, $\phi_c \approx 0.5$ at the surface. 
Visual inspection revealed a persistent coarse particle surface stripe oriented along the flow direction, akin to the surface pattern observed by~\citet{dOrtona2020} for broad inclined chute flows with $\alpha = 2$.
Second, for $\alpha = 6$ and 8 and $f_c = 0.1$ (for both $\alpha$) and 0.3 (for $\alpha = 8$), the profiles reveal a switch from usual to reverse segregation.
Thus, our results for $f_c = 0.1$ and 0.9 are in accord with the experiments of~\citet{thomas2000}. 
We examine the transition between usual and reverse segregation behaviors for these flows more thoroughly in Sec.~\ref{sec:trans}.
Note that results for $0.5 < f_c < 0.9$ are omitted from {Fig.~\ref{fig:fig4}} (and later figures); as we discuss in {Sec.~\ref{sec:velocity}}, this range for $\alpha = 6$ and 8 exhibit intermittent, disruptive bursts of activity that preclude a straightforward determination of the steady state.
However, between the bursts, the structure of these systems qualitatively recovers the layering and segregation behaviors depicted in {Fig.~\ref{fig:fig2}}.

In Fig.~\ref{fig:fig4}, oscillations in $\phi_c$ emerge for $\alpha = 4$ and higher for most $f_c$, corresponding to the layered structures visible in Fig.~\ref{fig:fig2}.
The peaks of the oscillations in $\phi_c$ never reach unity in Fig.~\ref{fig:fig4}, signifying that layers that are comparatively rich in coarse particles, i.e., rich compared to the regions of low $\phi_c$ that the oscillation minima entail, always contain an abundance of fines, which differentiates the interior layers from the surface block of coarse particles that forms under usual segregation.
In contrast, fine-rich layers are either devoid of coarse particles or are nearly so, as the minima in $\phi_c$ approach zero for $f_c \leq 0.5$, especially near the base.
Comparison of the $\phi_c$ oscillation spacing across different $\alpha$ indicates that the structure alternates between coarse-rich and fine-rich layers with an interval at least $(1+\alpha)d$, which is corroborated by the images in Fig.~\ref{fig:fig2}.
The height of the first $\phi_c$ peak for $f_c \leq 0.5$ is consistently shifted upwards by $\gtrsim \alpha d/2$ compared to the first peak for $f_c = 0.9$, suggesting that coarse particles may avoid interactions with the base by intervening planes of fines in fine-dominated scenarios.
As a final observation, while the lowest lying $\phi_c$ peaks for $f_c \leq 0.5$ tend to spatially coincide, by $z/H \sim 0.5$ the oscillations are sometimes out of sync, indicating that the oscillation wavelength depends weakly on the pile composition.

\begin{figure}
\centering
\includegraphics[width=0.45\textwidth]{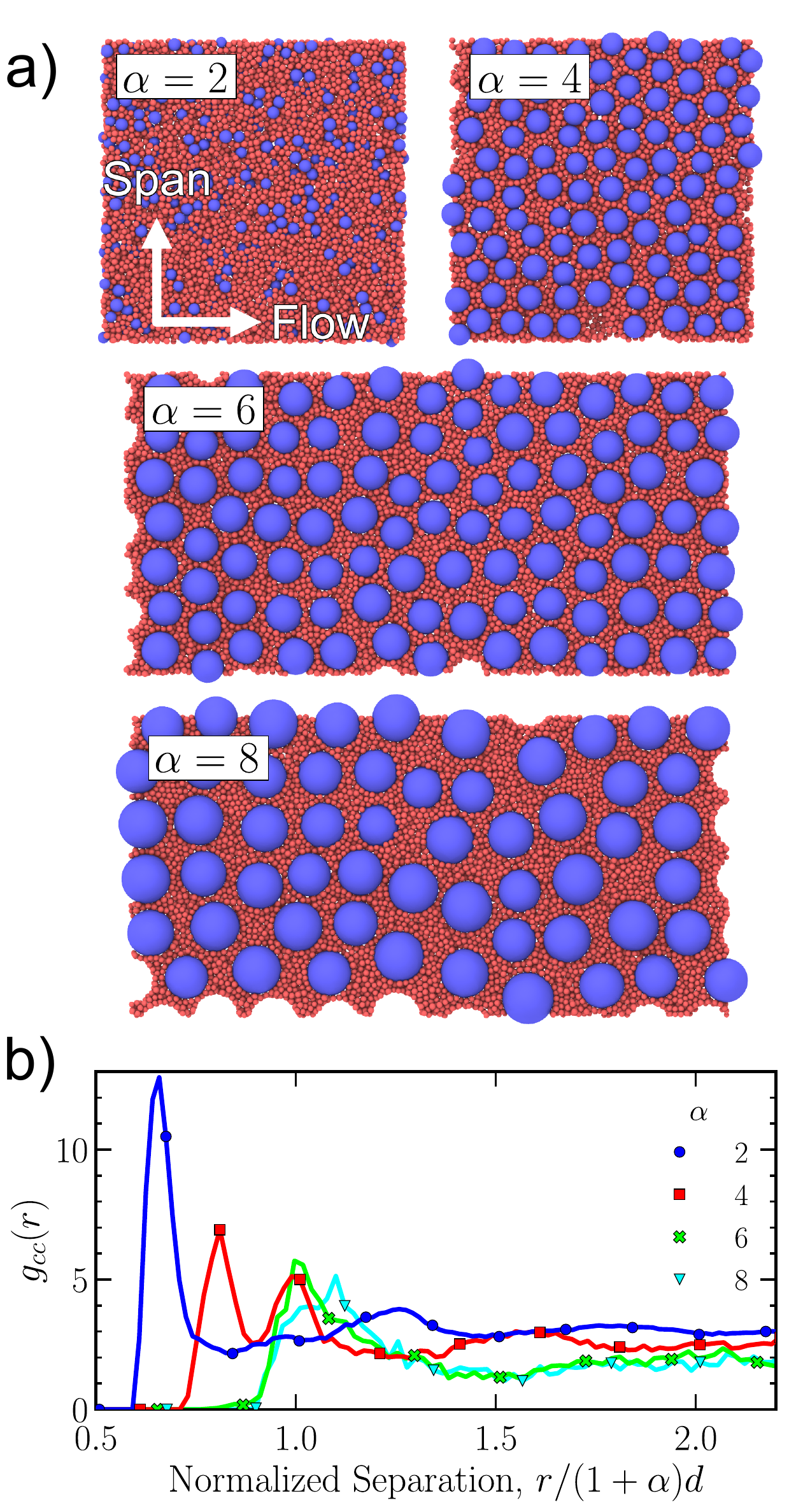}
\caption{
a) Top-down views of horizontal slices of depth $3d$ taken at $z/H\approx 0.5$ for the $f_c = 0.5$ snapshots depicted in Fig.~\ref{fig:fig2}a).
b) Three dimensional coarse-coarse radial distribution functions computed for the same flows as in a).
Markers are shown at fixed intervals to differentiate the curves.
}
\label{fig:fig5}
\end{figure}

The coarse particle concentration $\phi_c$ does not provide insight into the intra-layer particle arrangement.
To uncover the typical configuration of coarse-rich layers, Fig.~\ref{fig:fig5}a) shows top-down snapshots of horizontal slices taken at medium depth through the $f_c = 0.5$ piles shown in Fig.~\ref{fig:fig2}a).
These cross-sectional images reveal that coarse particles are rarely in direct contact---apart from in the coarse-dominant surface regions---despite commonly reaching $\phi_c > 0.5$ in the bulk for $\alpha = 4$ and higher.
Rather, coarse particles are usually separated by one or more fines and are often aligned in particle trains oriented along the flow direction~\mbox{\cite{fraysse2024}}.
Figure~\ref{fig:fig5}b) shows three dimensional radial distribution functions (RDFs) computed for pairs of coarse particles, $g_{cc}(r)$, considering every coarse particle in the pile.
While the pile structure is transverse isotropic and the RDF is an isotropic quantity, the RDFs nevertheless provide useful information about the local arrangement of coarse particles.
The inter-particle radial distance is normalized by $(1+\alpha)d$ to highlight the role of fines in limiting contact between coarse particles for large $\alpha$.
RDF results for both $\alpha = 2$ and 4 exhibit a contact peak owing to the sizable surface segregated domains for $f_c = 0.5$; however, the $\alpha = 4$ data also present an appreciable secondary peak at $r = (1+\alpha)d$.
The primary peak for $\alpha = 6$ corresponds to the same coarse particle separation, and the $\alpha = 8$ peak encompasses both $(1+\alpha)d$ and ($2+\alpha)d$, seemingly in agreement with the perceptibly larger intra-layer gaps between coarse particles in the corresponding image in Fig.~\ref{fig:fig5}a). 
This self-organized, alternating arrangement of coarse-fine-coarse lends itself to the hypothesis that fines facilitate flow by reducing the abundance of high friction contacts between coarse particles, which are associated with increased bulk friction~\cite{staron2016}.
It is interesting to note that the subsurface coarse particle ordering in the pile bulk occasionally encourages epitaxial ordering in the coarse-dominant region atop the flow (for example, for $\alpha = 4$ in Fig.~\ref{fig:fig2}).
The layer stacking in such a configuration is somewhat irregular, but roughly follows an A-B-A pattern.

\begin{figure*}
\centering
\includegraphics[width=0.95\textwidth]{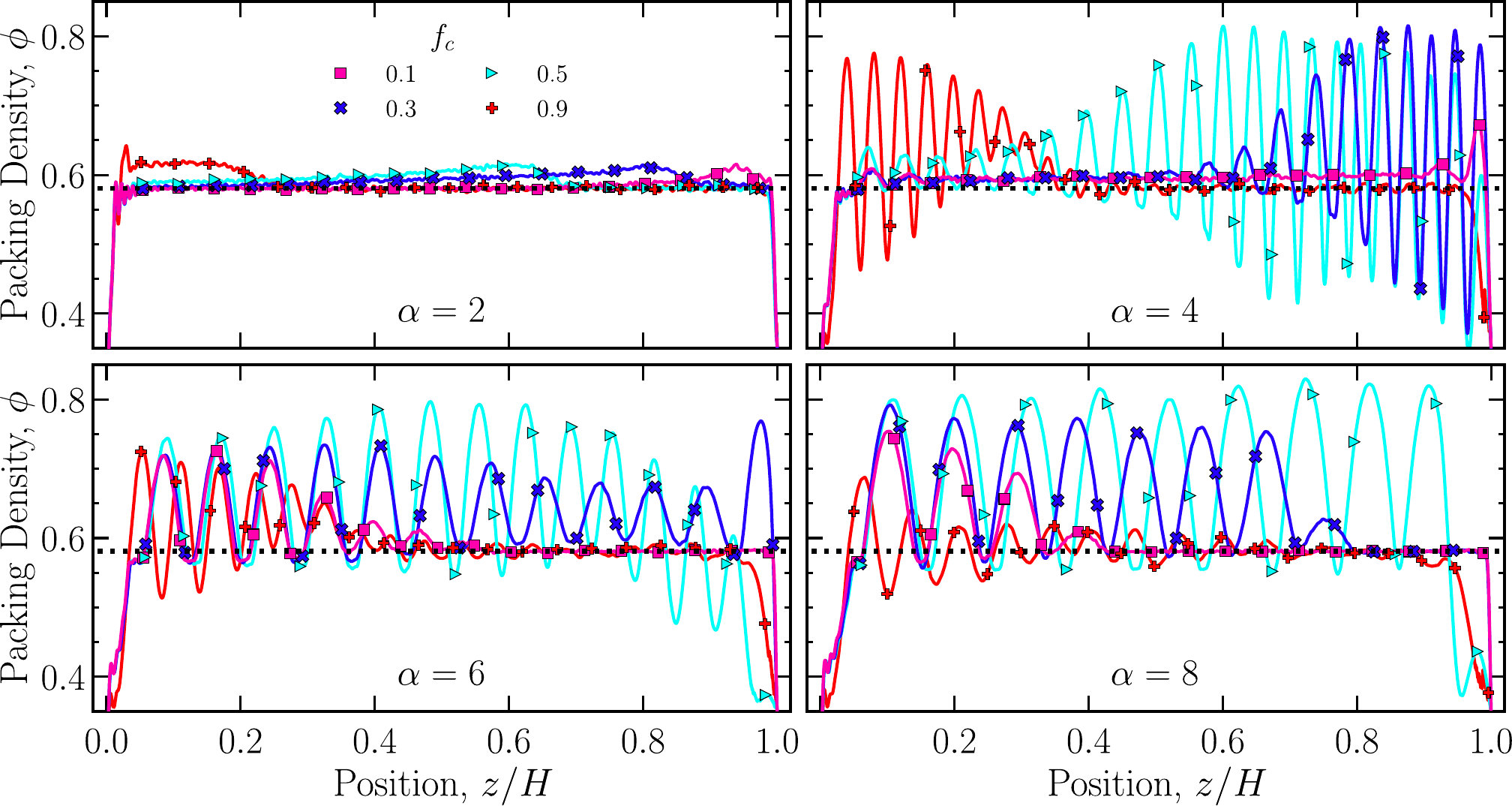}
\caption{
Steady state packing densities for each $\alpha$ and the selected values of $f_c$ indicated in the legend of the $\alpha = 2$ panel.
Horizontal dotted lines indicate $\phi_0 \approx 0.581$, obtained for a monodisperse system flowing under identical chute conditions.
Markers are shown at fixed intervals to differentiate the curves.
}
\label{fig:fig6}
\end{figure*}

A key consequence of the layered structure is that the bulk packing density, $\phi$, varies substantially between fine- and coarse-rich layers.
Figure~\ref{fig:fig6} shows $\phi$ profiles for the same $f_c$ as in Fig.~\ref{fig:fig4} and includes for comparison the mean packing density of an equivalent monodisperse chute flow, $\phi_0$, which gives $\phi_0 \approx 0.581$.
Single constituent free surface or basal domains composed of either fine or coarse particles maintain a comparable flow density to the monodisperse baseline for all $\alpha$ (c.f. $f_c = 0.1$ and 0.9), but both homogeneously mixed and layered regions reach higher $\phi$ for the same geometric reason that enables bidisperse assemblies to pack more efficiently~\cite{furnas1931,srivastava2021a}.
The increase in $\phi$ for $\alpha = 2$ is comparatively minor and varies smoothly with depth, mirroring the evolution of the corresponding $\phi_c$ shown in Fig.~\ref{fig:fig4}.
Conversely, $\phi$ reaches 0.8 or higher for the layered piles for $\alpha \geq 4$, and the peaks in $\phi$ exactly coincide with peaks in $\phi_c$.
Based on the analogy to jammed packings, we expected that peak $\phi$ values associated with coarse-rich layers would increase with increasing $\alpha$, yet taking $f_c = 0.5$, for instance, Fig.~\ref{fig:fig6} instead shows that the maximum $\phi$ achieved does not appear to systematically increase with $\alpha$.
Note that the local $\phi$ minima for $\alpha = 4$ dip substantially below $\phi_0$, while those for $\alpha = 6$ and 8 are usually close to $\phi_0$.
In combination with the earlier $\phi_c$ results and recognizing that $\alpha = 4$ represents a scenario in which fine particles fit poorly in between coarse particles~\mbox{\cite{gao2023}}, this finding implies that filling of gaps between coarse-rich layers is incomplete for $\alpha = 4$.

\begin{figure*}
\centering
\includegraphics[width=0.95\textwidth]{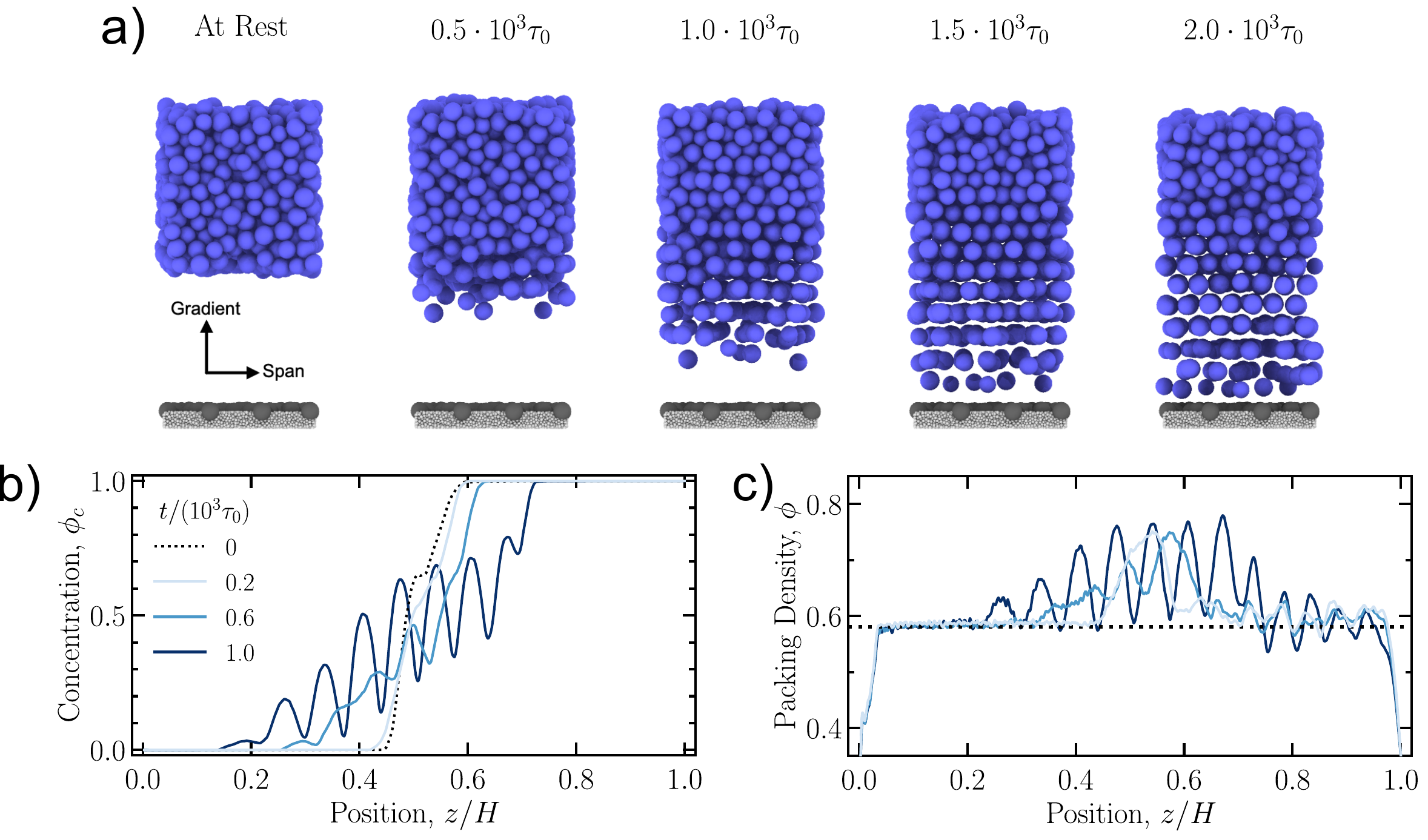}
\caption{
a) Snapshot sequence showing early time layer formation for $\alpha = 6$ and $f_c = 0.5$.
Flow is out of the page. 
Mobile fine particles have been removed to emphasize layering of coarse particles.
b) Coarse particle concentration profiles corresponding to the first three images in a).
c) Packing density profiles corresponding to the data in b).
The dotted line indicates $\phi_0 \approx 0.581$.
}
\label{fig:fig7}
\end{figure*}

To conclude this section, we elaborate on the origin of coarse- and fine-rich particle layers.
Figure~\ref{fig:fig7}a) contains a time sequence of snapshots depicting coarse particles and the base for $\alpha = 6$ and $f_c = 0.5$, which exhibited typical layering behavior [see Fig.~\ref{fig:fig2}a)].
The image sequence illustrates that the first layers form at the interface between the initially stacked coarse and fine particle domains.
When flow begins, some coarse particles migrate downward and exchange places with fines that have been propelled upward, producing a mixed zone at $t = 0.5\cdot 10^3\tau_0$.
By $t = 10^3\tau_0$, several layers are visible within the mixed zone, before any coarse particle encounters the base.
As time advances to $t = 1.5\cdot 10^3\tau_0$ and beyond, the alternating layered structure is largely formed and stabilized, after which it remains essentially unchanged.

Figure~\ref{fig:fig7}b) and c) show the associated evolution of coarse particle concentration and particle packing density over the first $\Delta t=10^3\tau_0$.
Profiles of $\phi_c$ indicate that weak ordering is already present at $t = 0.6\cdot 10^3\tau_0$ and by $t = 10^3\tau_0$, the layers are spread over a substantial fraction of the pile thickness.
The packing density profiles show an increase in $\phi$ that strengthens as the layers take shape.
Note that the peak $\phi$ values obtained at $t = 10^3\tau_0$ are nearly as high as the maximum values attained in the steady state.

The mechanism by which coarse particles descend through the pile appears to depend on $\alpha$, $f_c$, and secondarily, on the relative length of the span dimension.
The image sequence in Fig.~\ref{fig:fig7}a) shows an example of a configuration that is homogeneous along its span, which was the norm in our simulations (by design).
However, there were exceptions: for example, over the early flow stages for $\alpha = 4$ and $f_c = 0.1$, every coarse particle assembled into either the flow-oriented surface stripe described above or into a singular droplet shape just below the surface stripe.
Neither feature fully extended across the span dimension, and remarkably, the droplet was internally layered.
Eventually, the subsurface coarse particle structure was disrupted and devolved into a homogeneous distribution spread throughout the pile bulk (see Fig.~\ref{fig:fig4}), while the surface stripe remained.
Although it is outside the scope of the present work, we also observed vortex-like formation in separate simulations that considered $\alpha = 4$ systems with span dimensions greater than the pile height, reminiscent of the Rayleigh-Taylor instabilities reported by~\citet{dOrtona2020} for $\alpha = 2$.
These observations imply that the presence of secondary flows likely plays an important role in the ultimate segregation behavior for real systems, especially at early times for piles that are constructed far from their steady-state configuration. 

\subsection{Transition from Usual to Reverse Segregation}
\label{sec:trans}

\begin{figure*}
\centering
\includegraphics[width=0.9\textwidth]{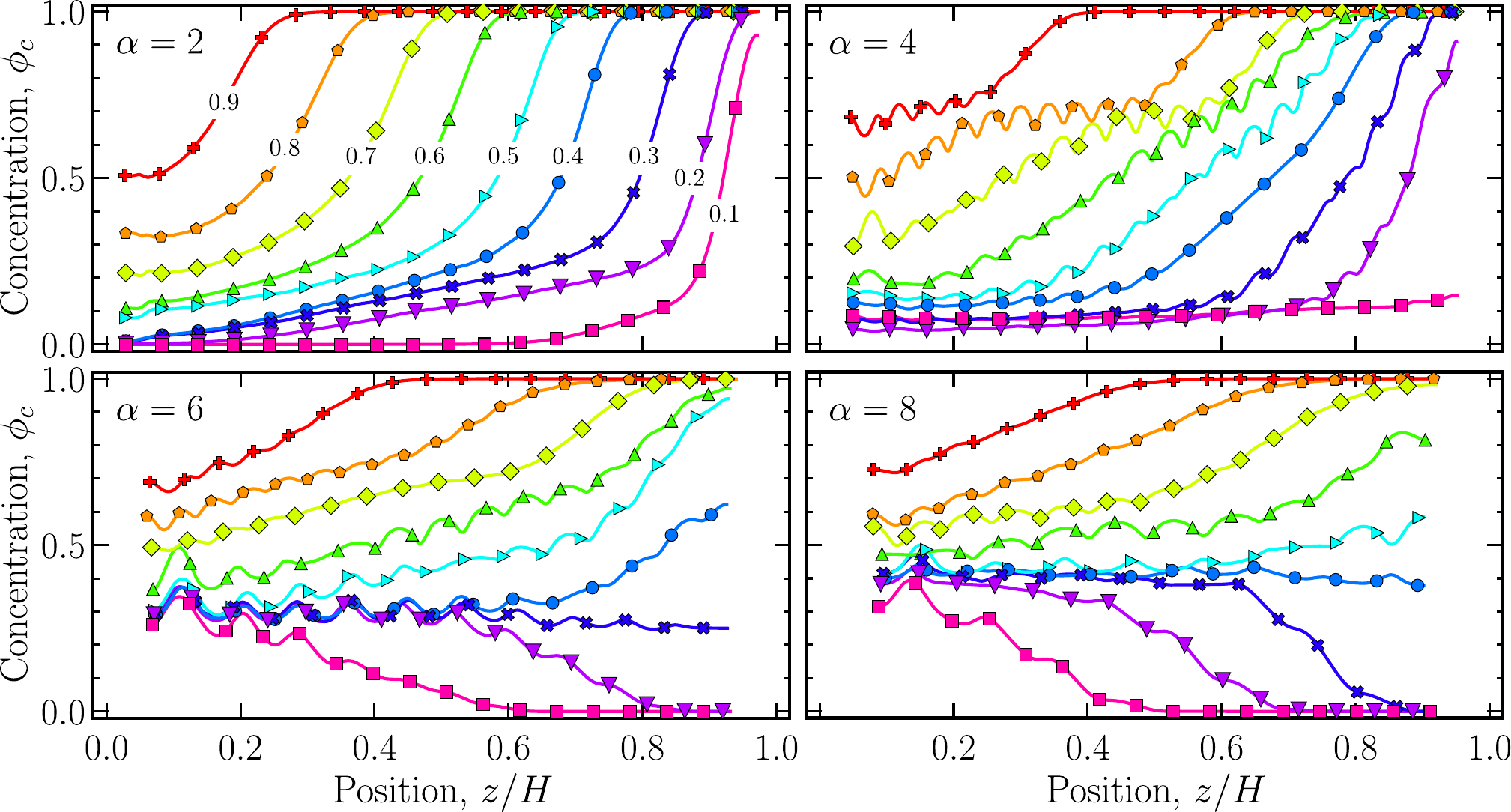}
\caption{
Steady state window-averaged coarse particle concentrations for the values of $f_c$ indicated in the legend of the $\alpha = 2$ panel.
The width of the averaging window is $2d(1+\alpha)$.
Markers are shown at fixed intervals to differentiate the curves.
}
\label{fig:fig8}
\end{figure*}

The concentration oscillations engendered by layer formation make it a challenge to compare profiles across all $f_c$ simultaneously and would likely prove irreconcilable for continuum models of segregation~\cite{gray2018,umbanhowar2019,gamble2025}.
To clarify general $\phi_c$ trends across the depth, we computed the rolling mean over a window of width $2d(1+\alpha)$ for each time-averaged data set~\cite{weinhart2012}.
The window-averaged results for all simulated combinations of $\alpha$ and $f_c$ are shown in Fig.~\ref{fig:fig8}.
Profiles for $\alpha = 2$ are minimally affected by the window because the variation of $\phi_c$ is smooth at the outset.
The notable aspects of the $\alpha = 2$ results are the increasing thickness of the stable segregated coarse particle domain as $f_c$ increases and the consistency of the concentration variation through the coarse-to-fine mixing zone.
After window-averaging, results for $\alpha = 4$ show similar behaviors to $\alpha = 2$ for $f_c > 0.1$: with oscillations suppressed, the concentration profiles exhibit a smooth rise from low $\phi_c$ at the base to unity in the coarse-dominant region at the pile surface, i.e., usual segregation.
However, the $f_c = 0.1$ profile stands apart by remaining nearly flat over the entire thickness.
This result is compatible with the observation of transitional behavior by~\citet{thomas2000} for inclined chute flows with $\alpha = 4.3$ and $f_c = 0.1$, where coarse particles were interspersed throughout the settled deposit and at the surface. 

Trends for $\alpha = 6$ and 8 are similar to those for $\alpha = 4$ in most respects---both include a nearly constant $\phi_c$ profile, but at different $f_c$: $f_c \approx 0.3$ for $\alpha = 6$ [see also Fig.~\ref{fig:fig2}b)] and $f_c \approx 0.4$ for $\alpha = 8$.
The salient finding of this analysis is that configurations with $f_c$ smaller than these special, uniform-across-the-depth configurations exhibit reverse segregation, while larger $f_c$ configurations undergo usual segregation.
Considering the concavity of the profiles in Fig.~\ref{fig:fig8}, the usual segregation curves have positive concavity with respect to $z$ and reverse segregation profiles have the opposite.
Most strikingly, the $f_c$ value for which the coarse-grained $\phi_c$ is spatially homogeneous, denoted $f_c^*$ for simplicity, increases with $\alpha$ from $f_c^*\approx 0.1$ for $\alpha = 4$ to $f_c^*\approx 0.4$ for $\alpha = 8$.
Similarly, the corresponding mean coarse particle concentration, $\phi_c^*$, for these special configurations also increases steadily with $\alpha$ from $\phi_c^* \approx 0.1$ for $\alpha = 4$ to $\phi_c^* \approx 0.4$ for $\alpha = 8$, seemingly representing an $\alpha$-dependent ideal concentration that is, perhaps coincidentally, close to $f_c^*$.
The overall segregation behaviors shown in Fig.~\ref{fig:fig8} are entirely consistent with the experimental measurements of~\citet{thomas2000}, who found reverse segregation in inclined flows with $f_c = 0.1$ and $\alpha \gtrsim 5$ and usual segregation for $f_c = 0.9$ for all examined $\alpha$ ($1.75 \lesssim \alpha \lesssim 10$).
In addition, the transition with increasing $f_c$ resembles tendencies observed in experiments on heaps and rotating drum flows~\mbox{\cite{thomas2000}}.

There are two final observations to derive from Fig.~\ref{fig:fig8}.
The first regards the upward mobility of fine particles during usual segregation.
In the initial resting pile, the interface between fine and coarse particle domains is located at $z/H = 1 - f_c$.
Considering results for $f_c = 0.9$ in Fig.~\ref{fig:fig8}, for instance, the steady-state profiles for each $\alpha$ show that the mixed zone ($0 < \phi_c < 1$) not only climbs much higher than its initial location, but also rises with increasing $\alpha$.
The elevation of fines higher within the pile occurs because, outside of a few edge cases, coarse particles exist at finite concentrations from the base to the surface.
Some fines are displaced upward as a consequence.
In addition, for high $f_c$ and to an increasing extent with increasing $\alpha$, there is always a dilute collection of fines that are rising or falling through the upper portion of the mixed zone under the dual influences of collisions and gravity, in what appears to be an equilibrium between the two processes.
The second observation from Fig.~\ref{fig:fig8} is that $\phi_c$ profiles for usual and reverse segregated piles are not mirrored for high $\alpha$.
The tendency displayed for $f_c\rightarrow 1$ is gradual growth toward $\phi_c = 1$ with $z$, whereas for $f_c\rightarrow 0$, $\phi_c \approx \phi_c^*$ until a height that increases with $f_c$, above which the decline to $\phi_c = 0$ is comparatively sharp.

\begin{figure*}
\centering
\includegraphics[width=0.9\textwidth]{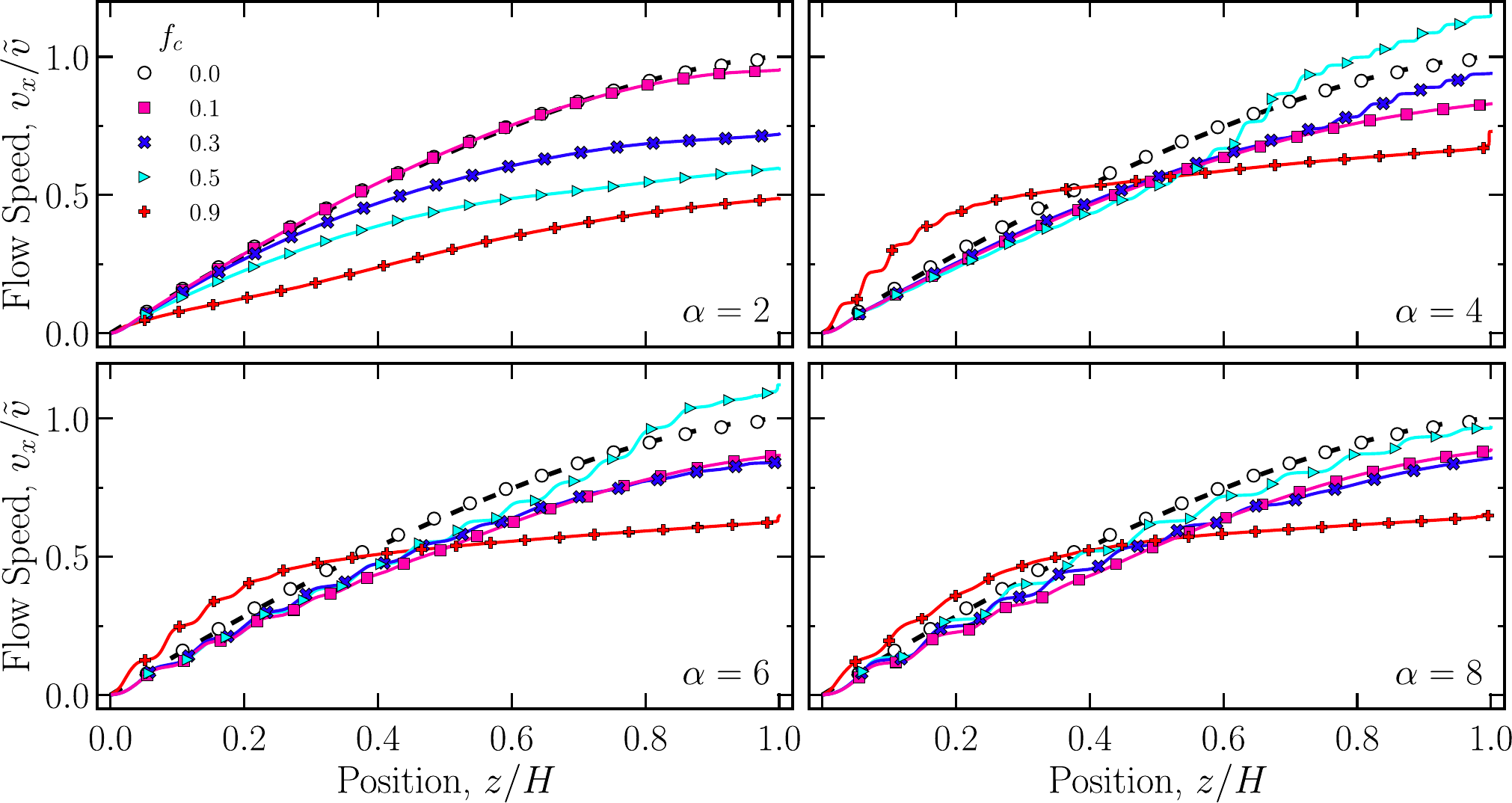}
\caption{
Steady state flow speed profiles for each $\alpha$ and the selected values of $f_c$ indicated in the legend of the $\alpha = 2$ panel.
Dashed lines indicate Eq.~\eqref{eq:vx} with normalization $\tilde{v}\equiv \left(2A_{\rm Bag}(\theta)/3\right)\sqrt{\phi_0 gH^3\sin{\theta}}$, where $A_{\rm Bag}(\theta = 22\degree) \approx 0.1 d^{-1}$.
Monodisperse data ($f_c = 0$) are reproduced in each panel. 
Markers are shown at fixed intervals to differentiate the curves.
}
\label{fig:fig9}
\end{figure*}

\begin{figure*}
\centering
\includegraphics[width=0.99\textwidth]{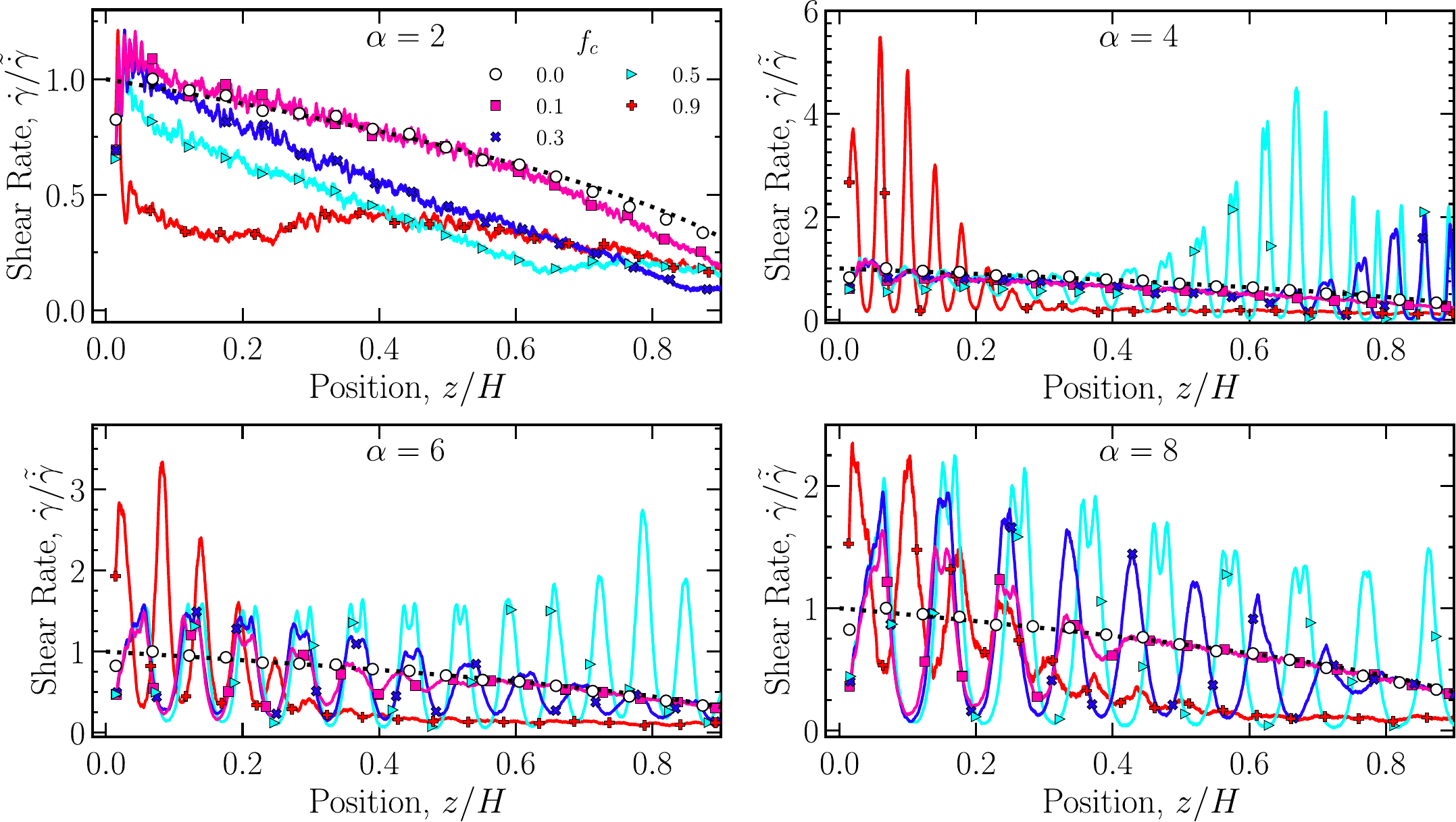}
\caption{
Shear rate profiles corresponding to the time-averaged $v_x$ profiles in Fig.~\ref{fig:fig9}.
The top $10\%$ of the flowing pile is omitted to highlight bulk behavior.
Dotted lines indicate $d v_x/dz$ evaluated for Eq.~\eqref{eq:vx} with normalization $\tilde{\dot{\gamma}}\equiv A_{\rm Bag}(\theta)\sqrt{\phi_0 gH\sin{\theta}}$.
Monodisperse data ($f_c = 0$) are reproduced in each panel. 
Markers are shown at fixed intervals to differentiate the curves.
}
\label{fig:fig10}
\end{figure*}

\subsection{Velocity and Shear Rate Profiles}
\label{sec:velocity}

Results in this section focus on comparisons between analytic expressions pertaining to monodisperse chute flows~\cite{silbert2001,midi2004} and flow profiles we obtained for bidisperse mixtures.
Throughout, we also include results for an identically treated fine particle monodisperse system ($f_c = 0$).
Note that no spatial averaging was performed for the results in this section.

Deep, stable chute flows of monodisperse particles exhibit flow speed profiles that obey the Bagnold velocity scaling $v_x \propto \left[ H-(H-z)^{3/2} \right]$~\cite{bagnold1954,silbert2001,midi2004}.
From~\citet{silbert2001}, the full expression is
\begin{equation}
    v_x(z)=A_{\rm Bag}(\theta)\frac{2}{3}\sqrt{\phi_0 gH^3\sin{\theta}}\left[1-\left(1-\frac{z}{H}\right)^{3/2}\right],
    \label{eq:vx}
\end{equation}
where $A_{\rm Bag}(\theta)$ depends weakly on the particle contact model~\cite{silbert2001}.
In steady state, $v_y$ and $v_z$ vanish on average in the absence of secondary flows, which we intentionally suppressed by constructing a tall, narrow pile~\cite{dOrtona2020}.
It is worthwhile to keep in mind that slip between the flowing pile and the base depends on both base construction and pile composition: smooth bases and those composed exclusively of fines, regularly arranged or otherwise, promote slip of coarse particles~\cite{silbert2001,silbert2002,rognon2007,weinhart2012,kumaran2012}, which manifests in $v_x$ profiles as a lateral shift.
Since we chose to create the roughened base using both fine and coarse particles, basal slip was minimized in our simulations.

Figure~\ref{fig:fig9} shows $v_x$ profiles across different $\alpha$ and $f_c$, normalized by the monodisperse surface value $\tilde{v}\equiv \left(2A_{\rm Bag}(\theta)/3\right)\sqrt{\phi_0 g H^3 \sin \theta}$, where $A_{\rm Bag}(\theta)$ was determined by a least squares fit.
We found excellent agreement between our monodisperse system and Eq.~\eqref{eq:vx} with $A_{\rm Bag}(\theta = 22\degree) \approx 0.1 d^{-1}$, in quantitative agreement with values reported by~\citet{silbert2001}.
In keeping with findings by other authors~\cite{rognon2007,tripathi2011,staron2016,barker2021}, results for the modest bidispersity entailed by $\alpha = 2$ show that progressively increasing the coarse particle mass fraction leads to slower flow speeds that increasingly depart from the $f_c = 0$ profile.
For $\alpha \geq 4$, the trends in Fig.~\ref{fig:fig9} with $f_c$ become less clear, but are generally consistent across $\alpha$. 
Focusing on $f_c < 0.5$, increasing $f_c$ reduces the surface flow speed below $\tilde{v}$, but the overall profile shape is maintained.
In the Supplemental Materials, we show that Eq.~\eqref{eq:vx} continues to adequately describe the data for $f_c \lesssim 0.4$ for $\alpha = 4-8$, with only a 10\%--20\% reduction in $A_{\rm Bag}(\theta)$ values obtained from fitting (depending mainly on $f_c$).
It appears that $f_c = 0.5$ is a borderline case for $\alpha \geq 4$ because the $\alpha = 4$ and 6 results differ from the Bagnold profile but the $\alpha = 8$ results conform. 
In contrast, $f_c \gtrsim 0.6$ flow profiles are not well described by Eq.~\eqref{eq:vx} (see the Supplemental Materials).
For $f_c = 0.9$, specifically, flow is faster than the Bagnold profile over the bottom $\sim 40\%$ of the pile, where fine and coarse particles coexist, and slower than the Bagnold profile over the remainder of the thickness, which is entirely composed of coarse particles.

Beginning with $\alpha = 4$, the impact of particle layer formation on $v_x$ is perceivable in Fig.~\ref{fig:fig9}.
Within each coarse-rich layer of width $\sim\alpha d$, $v_x$ is necessarily uniform because the alternative would produce in-plane gaps that would presumably be filled by rising or falling fines from adjacent fine-rich layers.
Figure~\ref{fig:fig9} reveals that $v_x$ increases rapidly between coarse-rich layers, resulting in stair-stepped profiles (albeit somewhat smoothed out by time averaging and partial layer mixing).
Such non-smooth speed profiles have been observed previously for shallow monodisperse chute flows~\mbox{\cite{weinhart2012}} and in (two dimensional) chute flows of large bidispersity (see Fig. 2 of ~\citet{rognon2007}).

The local shear rate, $\dot{\gamma} \equiv dv_x/dz$, is small or vanishes entirely in coarse-rich layers, while in fine-rich layers it is higher than the value obtained at matching depth in the equivalent $f_c = 0$ flow.
Figure~\ref{fig:fig10} shows the normalized shear rate for the flow speed profiles in Fig.~\ref{fig:fig9}.
Shear rate results for $\alpha = 2$ quantify the braking effect of increasing $f_c$ and hint at non-Bagnold dynamics that the raw $v_x$ profiles belie.
For $\alpha = 4$ and higher, computed shear rates reveal how sharply layering modifies the flow profile---as expected, the shear rate spikes between coarse particle layers and drops to small values inside the layers.
Some of the peaks are bifurcated (e.g., $f_c = 0.5$), suggesting that fine-rich layers with thicknesses of multiple $d$ may have their own internal structure.
Interestingly, shear rates for $f_c = 0.1$ usually adhere to the predicted profiles in the fine-dominant region, irrespective of $\alpha$. 
Under usual segregation for $\alpha = 2$, the shear rate is nominal near the base and reduced near the surface. 
For reverse segregation for $\alpha \geq 6$, it is the pile surface that matches the Bagnold profile, despite the dynamics of the mixed region near the base.

For systems in the range $0.5 < f_c < 0.9$ for $\alpha \geq 4$, the time-averaged $v_x$ profiles exhibited altered behavior and faster flow at the surface, which as we have noted led us to omit the data from Fig.~\ref{fig:fig9}.
The associated results are included in the Supplemental Materials, but there was a curious behavior in some scenarios that bears mentioning.
During several simulations for $\alpha = 6$ and 8, we observed an intermittent flow instability that manifested as periodic bursts of activity that rapidly swelled the pile and were associated with temporarily increased kinetic energy.
In terms of the $v_x$ profiles, the intermittency produced dramatically increased mean surface flow speeds and, in some cases, accelerated the bulk flow as well.
Mechanistically, the heightened activity appeared to correspond to the disruption of alternating, ordered fine and coarse particle layers akin to those shown in Fig.~\ref{fig:fig2}, which are stable, steady-state arrangements for $f_c \lesssim 0.5$ (and $f_c = 0.9$) but are apparently only metastable for certain values of $f_c$---particularly for $f_c$ that give the highest packing density~\cite{furnas1931,srivastava2021a}.
Similar ordering-disordering transitions, temporally-periodic or otherwise, have been observed in monodisperse systems, usually in connection with specific basal structures or basal particle sizes~\mbox{\cite{silbert2002,kumaran2012, weinhart2012,kumaran2013,yang2022}}.
~\mbox{\citet{yang2022}} suggested that the monodisperse order-to-disorder transition initiates at the surface and propagates through the depth of the pile as the bulk heats up (in a granular temperature sense); however, in our bidisperse systems, it appears that layer disruption anywhere in the pile, not just at the surface, is capable of triggering the intermittency.
A simple explanation in the bidisperse case is that breakdown of the layered structure forces the pile to locally dilate because layer formation enables a greater degree of compactness than disordered configurations can sustain---the dilation then rapidly spreads through the rest of the pile.
Further investigation into the origin of the transition in highly bidisperse systems is needed to conclusively ascertain the factors that cause it (e.g., base structure, incline angle, and pile depth~\mbox{\cite{weinhart2012}}).
We confirmed that intermittency did not depend on whether the base contained coarse particles, but we did not comprehensively study effects of base structure in this work.

\begin{figure}
\centering
\includegraphics[width=0.47\textwidth]{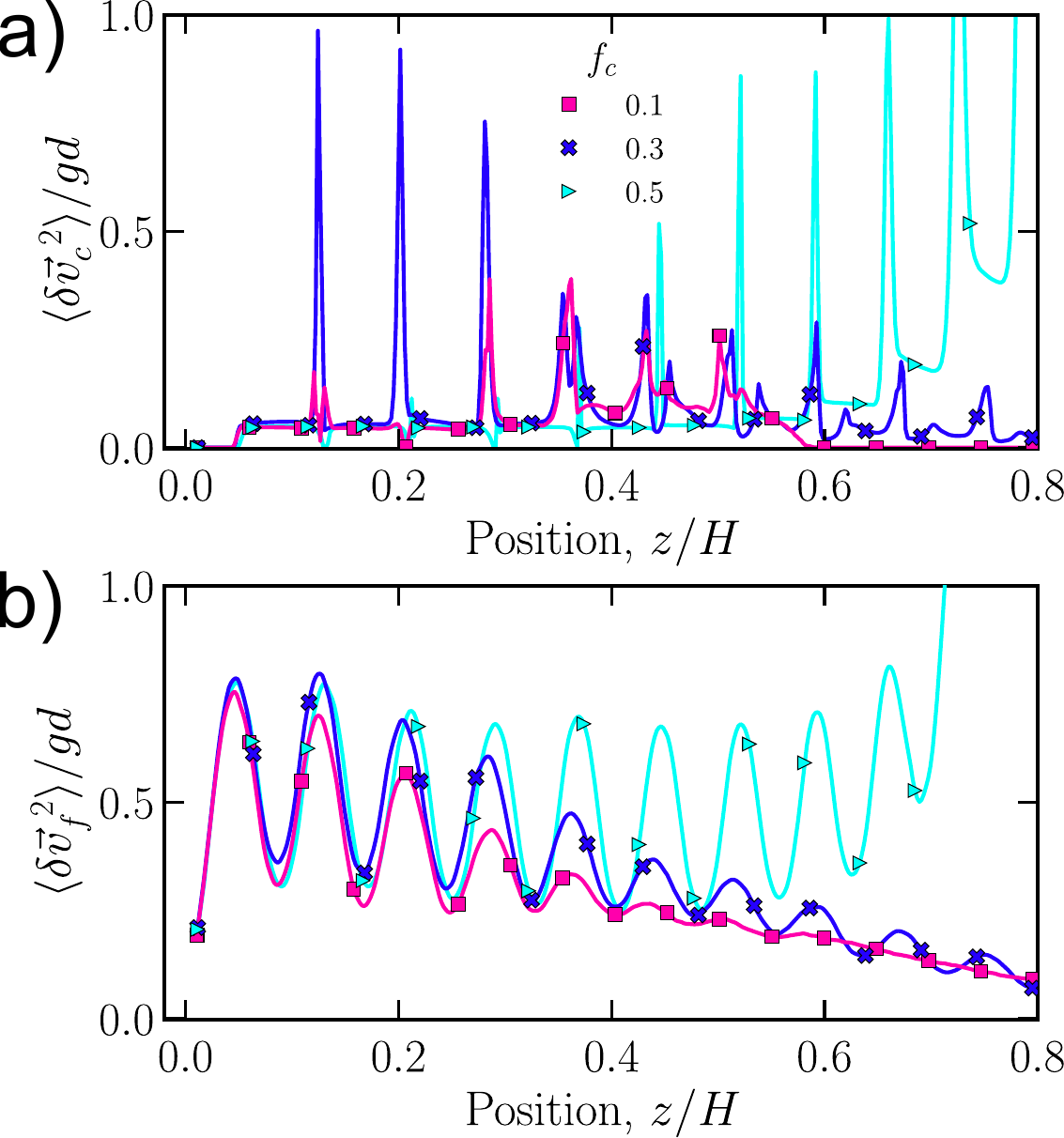}
\caption{
Steady state subsurface profiles of normalized velocity variance for $\alpha = 6$ over $z \leq 0.8H$ for coarse particles [a)] and fine particles [b)] for selected $f_c$.
Markers are shown at fixed intervals to differentiate the curves.
}
\label{fig:fig11}
\end{figure}

The steady-state layered particle arrangement and corresponding non-smooth flow speed profiles raise the question of what particle activity occurs in the bulk.
To deconvolve particle motion relative to the mean pile flow, we report velocity variance profiles partitioned between fine and coarse particles, $\langle \delta \vec{v}_f^{\ 2}\rangle$ and $\langle \delta \vec{v}_c^{\ 2}\rangle$, where $\langle\cdot\rangle$ refers to an average taken over each $z$-bin and $\delta \vec{v}_i = \vec{v}_i(z) - \langle\vec{v}_i(z)\rangle$ for $i = f$ or $c$.
Figure~\ref{fig:fig11}a) shows results for coarse particles for $\alpha = 6$ and selected $f_c \leq 0.5$, which are emblematic for $\alpha = 6$ and 8.
The figure shows profiles over the range $z \leq 0.8H$ to avoid surface effects.
Unsurprisingly, coarse particle velocity fluctuations are suppressed within coarse-rich layers with a low value that is approximately constant for the $f_c$ plotted in the figure.
Sharp spikes in the variance correspond to coarse particles that extend outside their layers and arise entirely from the flow direction component (which we examined separately).
Coarse particles that jut out of or escape from the principal layers tend to be moving at disparate speeds relative to other coarse particles intersecting the same plane (as well as to the fines above and below, as Fig.~\mbox{\ref{fig:fig9}} showed), yielding the observed increased variance.
Interesting avenues for future work pertaining to layer stability include addressing such questions as how strongly coarse particles are bound within their layers and how frequently they jump between layers, answers for which could provide rich insights into the drivers behind layering.

The velocity variance of fines is comparatively large in both coarse-rich and fine-rich layers, as shown in Fig.~\ref{fig:fig11}b).
Fine-rich layers exhibit the highest typical velocity variances.
The fluctuations of fines inhabiting coarse-rich layers are lower, but still exceed those of their coarse particle counterparts by a sizable factor. 
The reduction of $\langle \delta \vec{v}_f^{\ 2}\rangle$ in coarse-rich layers is enforced by the high intra-layer density and by the relative uniformity of the flow of the more massive coarse particles.
We verified that the individual velocity component contributions are comparable for fines, though the peak-to-valley range along the flow direction is greater than in the orthogonal directions.
Outside layered regions, Fig.~\ref{fig:fig11}b) shows that $\langle \delta \vec{v}_f^{\ 2}\rangle$ decays smoothly toward the pile surface, as can be seen for $f_c = 0.1$, which lacked coarse particles above the middle of the pile, and for $f_c = 0.3$, which marked the border between usual and reverse segregation for $\alpha = 6$ (as Fig.~\ref{fig:fig8} showed).

The partitioned velocity variances shown in Fig.~\ref{fig:fig11} suggest a possible explanation for the reinforcement of inter-species layering, if not for the layer formation process itself.
Coarse particles that protrude from existing coarse-rich layers experience a discrepancy in $v_x$ from neighbors above and below.
High relative velocity collisions thus restore the deviating coarse particle back to the layer plane.
Moreover, while fines are highly active in fine-rich layers, their migration to nearby, lower activity, coarse-rich layers is hindered by the lack of available space. 
A similar dearth of free space at large diameter ratios has been connected to increased segregation time scales in annular shear cell experiments~\mbox{\cite{golick2009}}.
Despite the variations of coarse-rich $\phi$ shown earlier in Fig.~\ref{fig:fig6}, the fact that the flow usually achieves steady state with respect to layering is indicative that a balance of fine particle migration between coarse- and fine-rich layers is eventually reached.
However, the special scenarios briefly described earlier in this section for which intermittent flow instabilities occurred, which temporarily disrupted the layered structure, signify that equilibrium cannot always be maintained.
Such intermittency and the conditions that bring it about remain an interesting avenue for future work.

\section{Conclusions}
\label{sec:conc}
We performed discrete element method simulations of inclined chute flow of bidisperse granular mixtures of otherwise identical spherical particles.
The degree of bidispersity simulated ranged from the conventional limit of coarse particles possessing mean diameters twice the diameter of the fine species to a maximum mean diameter ratio of eight.
The simulation geometry was constructed as a deep, narrow channel to deliberately suppress secondary flows.
Depending on the particle diameter ratio and the coarse particle mass fraction, the simulations revealed transitions from usual segregation---referring to the scenario in which the flowing pile surface was covered by coarse particles---to a situation of reverse segregation---in which all coarse particles were immersed in the pile.
For diameter ratios of four and higher, specific mass fractions at which window-averaged local coarse particle concentrations were spatially uniform were identified as threshold values.
Above the threshold, usual segregation prevailed; below, reverse segregation was observed. 
The simulations showed that the mass fraction threshold consistently increased with the degree of bidispersity over the range considered. 

A recurring observation for bidispersity of four and higher was the development of alternating coarse and fine particle layers shortly after flow began.
Rather than favoring true segregation vis-\'a-vis complete spatial separation, layered piles promoted local mixing by self-organizing into alternating horizontal planes rich in fine or coarse particles.
Fine-rich layers were nearly devoid of coarse particles, while coarse-rich layers displayed a range of concentrations that generally depended on the overall mass fraction and the layer position within the pile.
Fine particles within the coarse-rich layers were typically located between coarse particles, which enhanced the local packing density to values that commonly reached and exceeded 0.8, signifying a high degree of compactness.
Having the ability to control such layered structures could offer a complementary  approach to methods such as the epitaxial growth of granular packings \cite{panaitescu2014} in the design of granular packing construction with tailored structures.

Layer formation also affected flow speed profiles computed for the bidisperse piles.
In piles where the overall coarse particle abundance was small, the general effect of bidispersity was to reduce the prefactor that sets the surface and bulk flow speeds and to lower the local shear rate.
Layering introduced coincident shear rate oscillations---the shear rate was low within coarse-rich layers and jumped in fine-rich layers---and was associated with spatially varying particle velocity fluctuations. 
Reduced velocity fluctuations of ordered coarse particles correspondingly suppressed fluctuations for fine particles in coarse-rich layers, while fine-rich layers exhibited higher velocity fluctuations.
For high mass fractions of coarse particles, conversely, flow proceeded faster than an equivalent thickness monodisperse configuration over the lowest $\sim 40\%$ of the pile and slower than the monodisperse configuration over the rest of the thickness.
This result and the general structure that organically emerged in coarse-rich layers support the notion that fines have a lubricating influence that facilitates flow by eliminating direct coarse particle contact.
This finding hints that judicious construction of the particle size composition of the pile can have significant ramifications for improving its flowability, a nearly universal aim in granular materials research.
Furthermore, gaining a deeper understanding of segregation behavior in varied granular mixtures at the particle scale could lead to the development of more effective mixing or demixing techniques for implementation at commercial and industrial scales.

The results of this study are directly applicable for extending continuum segregation models of size-bidisperse flows to large particle diameter ratios~\mbox{\cite{gray2018,umbanhowar2019,barker2021,liu2023,gamble2025}}.
While pronounced layering at the particle scale is somewhat antithetical to continuum modeling, useful information can doubtless still be gleaned by coarse-graining over the oscillating profiles.
Looking farther forward, flows of particles distributed continuously over a broad range of sizes contain interactions between particles of small and large size disparity alike.
The large particle diameter ratios considered here therefore represent an important step toward investigating size distributions of geophysical relevance, which often span multiple, and sometimes many, orders of magnitude between smallest and largest particle sizes.

\section{Supplemental Materials}
Additional material showing steady-state pile surface heights, Bagnold profile comparisons and fit parameters, and flow speed profiles for $f_c \geq 0.5$ are included.

\section{Acknowledgements}
I.S. acknowledges support from the U.S. Department of Energy (DOE), Office of Science, Office of Advanced Scientific Computing Research, Applied Mathematics Program under Contract No. DE-AC02-05CH11231.
The authors acknowledge useful conversations with Nico Gray, Christopher Johnson, and Philip Gamble of the University of Manchester and with Paul Umbanhowar, Richard Lueptow, and Dhairya Vyas of Northwestern University.
This work was performed in part at the Center for Integrated Nanotechnologies, a U.S. DOE and Office of Basic Energy Sciences user facility.

Sandia National Laboratories is a multi-mission laboratory managed and operated by National Technology \& Engineering Solutions of Sandia, LLC (NTESS), a wholly owned subsidiary of Honeywell International Inc., for the U.S. Department of Energy’s National Nuclear Security Administration (DOE/NNSA) under contract DE-NA0003525. This written work is authored by an employee of NTESS. The employee, not NTESS, owns the right, title and interest in and to the written work and is responsible for its contents. Any subjective views or opinions that might be expressed in the written work do not necessarily represent the views of the U.S. Government. The publisher acknowledges that the U.S. Government retains a non-exclusive, paid-up, irrevocable, world-wide license to publish or reproduce the published form of this written work or allow others to do so, for U.S. Government purposes. The DOE will provide public access to results of federally sponsored research in accordance with the DOE Public Access Plan.

\bibliography{bibfile}

\widetext
\clearpage

\begin{center}
\textbf{\large Supplemental Materials: Reverse segregation and self-organization in inclined chute flows of bidisperse granular mixtures}
\end{center}

\setcounter{figure}{0}
\setcounter{section}{0}
\makeatletter
\renewcommand{\thefigure}{SM\arabic{figure}}

\section{Flowing pile surface height}

\begin{figure}[H]
\centering
\includegraphics[width=0.5\textwidth]{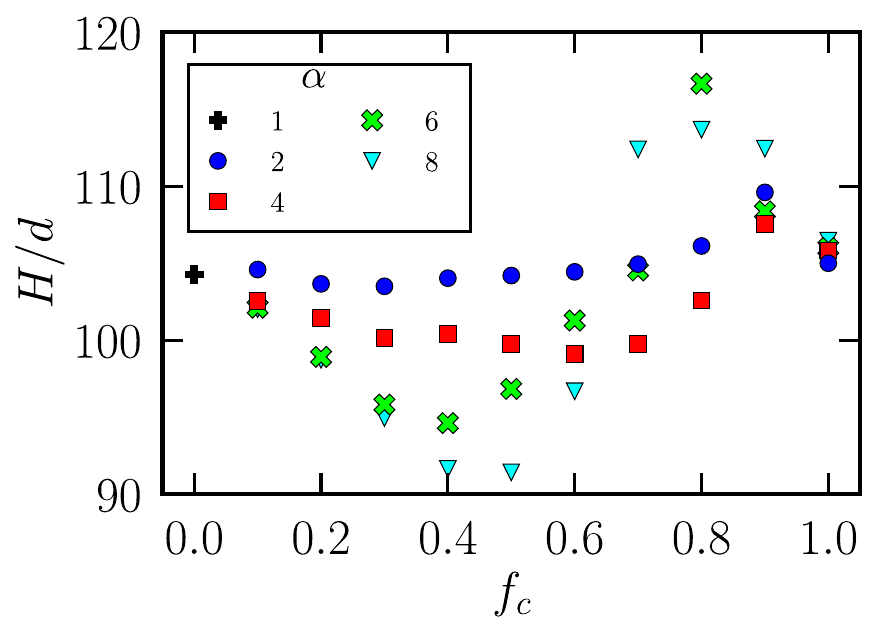}
\caption{
Surface height of flowing piles as a function of $f_c$ for the indicated $\alpha$ values.
}
\label{fig:SM1}
\end{figure}

Figure~\ref{fig:SM1} shows flowing pile height, $H$, defined as the highest position with particle volume fraction $\phi \geq 0.35$.
The $H$ values are obtained in steady-state, aside from those for $\alpha = 6$ and 8 and $f_c \sim 0.7 - 0.8$, which exhibited the intermittent disruptive behavior described in the main text.
The reduction in $H$ at intermediate $f_c$ occurs because bidisperse mixtures consolidate more efficiently as the size ratio increases.

\section{Bagnold velocity scaling comparison}

\begin{figure}[H]
\centering
\includegraphics[width=\textwidth]{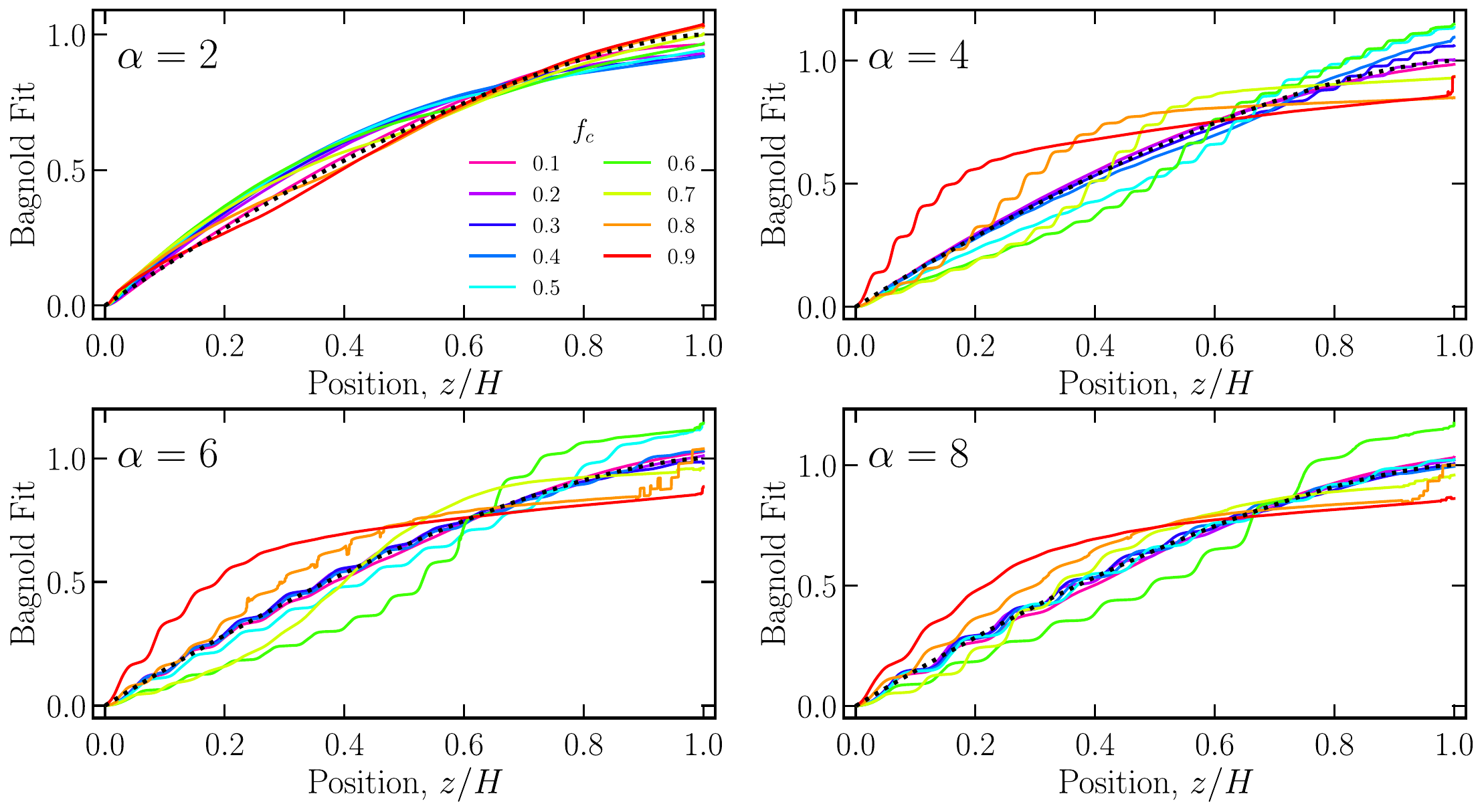}
\caption{
Time-averaged $v_x(z)$ profiles compared to the corresponding Bagnold fit for four $\alpha$ and the indicated $f_c$ values.
The dotted line corresponds to the Bagnold $v_x(z)$ profile.
}
\label{fig:SM2}
\end{figure}

\begin{figure}[H]
\centering
\includegraphics[width=0.5\textwidth]{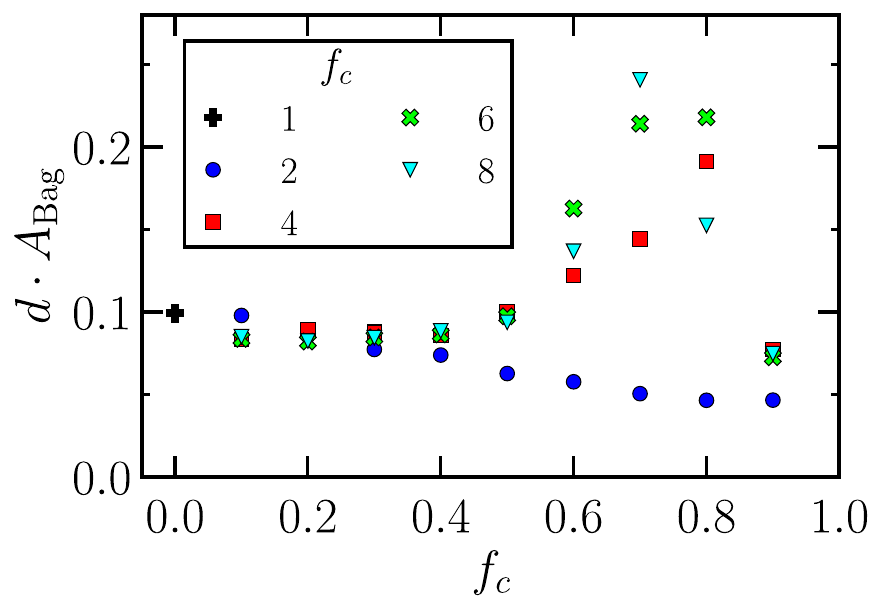}
\caption{
$A_{\rm Bag}\left(\theta = 22\degree\right)$ fit values as a function of $f_c$ for the indicated $\alpha$.
}
\label{fig:SM3}
\end{figure}

Figure~\ref{fig:SM2} shows comparisons between time-averaged, steady-state flow speeds normalized by prefactors obtained through least squares fits of the data to the Bagnold velocity scaling (setting $\phi_0$ = 0.581).
As plotted, Fig.~\ref{fig:SM2} highlights the discrepancy that arises between the Bagnold profile and the high $f_c$ results for $\alpha \geq 4$.
The corresponding fitted values of $A_{\rm Bag}(\theta = 22\degree)$ are shown in Fig.~\ref{fig:SM3}.
All else being equal, $\alpha d\cdot A_{\rm Bag}(\theta) = constant $ for monodisperse systems (i.e., $f_c = 0$ or $1$).
$A_{\rm Bag}(\theta = 22\degree)$ consistently decreases with increasing $f_c$ for $\alpha = 2$, while results for $\alpha \geq 4$ deviate from this trend for $f_c \geq 0.5$, underscoring the fundamental difference in flow behavior that occurs for coarse-particle-dominated systems with large size ratios.

\section{Flow speed profiles}

\begin{figure}[H]
\centering
\includegraphics[width=\textwidth]{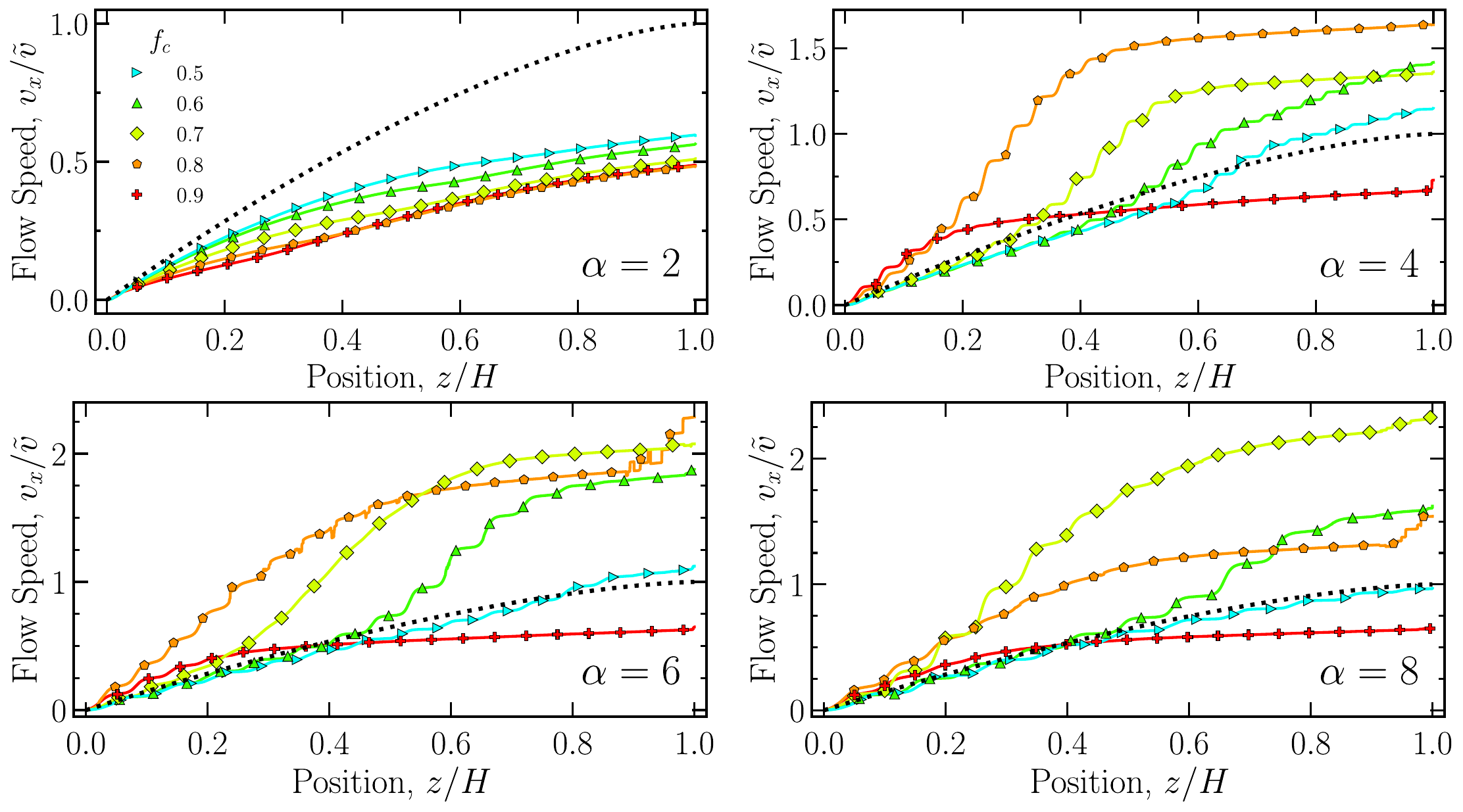}
\caption{
Time-averaged flow speed profiles for the indicated $f_c$ values. 
The dotted line corresponds to the Bagnold $v_x(z)$ profile.
}
\label{fig:SM4}
\end{figure}

Figure~\ref{fig:SM4} shows several additional time-averaged $v_x(z)$ profiles that were omitted from the main text (with $f_c = 0.5$ and 0.9 reproduced).
Given the flow intermittency that occurred for some of these systems, the results cannot be generally categorized as steady state.
For $\alpha \geq 4$, the pile surface uniformly flows at a greater speed than $\tilde{v}\equiv \left(2A_{\rm Bag}(\theta)/3\right)\sqrt{\phi_0 gH^3\sin{\theta}}$ for $0.6 \leq f_c \leq 0.8$, but the $f_c = 0.9$ pile surface always flows more slowly.
Stair-stepping in the $v_x(z)$ profiles is prominent for $f_c \geq 4$.

\end{document}